\documentclass[letterpaper,twocolumn,10pt]{article}
\usepackage{usenix2019_v3}

\usepackage[english]{babel}

\usepackage{epsfig}
\usepackage{balance}
\usepackage{pifont}
\usepackage[english]{babel}
\usepackage{times}
\usepackage{graphicx} 
\usepackage{subcaption}
\usepackage{tabularx}
\usepackage[noend]{algorithm2e}
\usepackage{xspace}
\usepackage{listings}
\usepackage{verbatim}
\usepackage{booktabs}
\usepackage{colortbl}
\usepackage{nicefrac}
\usepackage{siunitx}
\usepackage{stackengine}
\usepackage{titlesec}
\usepackage{titling}
\titlespacing*{\section}{0pt}{4pt}{3pt}
\titlespacing*{\subsection}{0pt}{4pt}{3pt}
\titlespacing*{\subsubsection}{0pt}{4pt}{3pt}
\usepackage{amsmath,amsthm,amssymb}
\usepackage{mathtools}

\usepackage[subtle]{savetrees}
\newtheorem{theorem}{Theorem}
\usepackage{color}
\definecolor{darkred}{rgb}{0.7,0,0}
\definecolor{darkgreen}{rgb}{0,0.5,0}
\hypersetup{colorlinks=true,
        linkcolor=darkred,
        citecolor=darkgreen}
\newcommand{\eg}{{e.g.,} }

\newcommand{\ie}{{i.e.}, }

\newcommand{\Fig}[1]{Fig.~\ref{fig:#1}\xspace}

\newcommand{\Sec}[1]{$\S$\ref{s:#1}\xspace}
\newcommand{\App}[1]{Appendix~\ref{app:#1}\xspace}

\newcommand{\ma}[1] {{\textcolor{blue}{MA: #1}}}
\newcommand{\pg}[1] {{\textcolor{red}{PG: #1}}}

\newcommand{\an}[1] {{\textcolor{blue}{AN: {#1}}}}

\newcommand{\cut}[1]{}

\newcommand{\name}{ABC\xspace}

\def\compactify{\itemsep=0pt \topsep=0pt \partopsep=0pt \parsep=0pt}
\let\latexusecounter=\usecounter

\newenvironment{CompactEnumerate}
  {\def\usecounter{\compactify\leftmargin=13pt\latexusecounter}
   \begin{enumerate}}
  {\end{enumerate}\let\usecounter=\latexusecounter}

\makeatletter
\patchcmd{\env@cases}{1.2}{0.72}{}{}
\makeatother

\begin{document}
\setlength{\droptitle}{-1.2cm}

\title{\Large \bf  ABC: A Simple Explicit Congestion Controller for Wireless Networks\vspace{-3mm}}

\author{Prateesh Goyal$^{1}$, Anup Agarwal$^{2}$\thanks{Work done largely while a visiting student at MIT CSAIL.}, Ravi Netravali$^{3}$\thanks{Work done largely while a PhD student at MIT CSAIL.}, Mohammad Alizadeh$^{1}$, Hari Balakrishnan$^{1}$\\
$^{1}$MIT CSAIL, $^{2}$CMU, $^{3}$UCLA}
\date{\vspace{-8mm}}
%
\maketitle
\begin{abstract}
We propose {\em Accel-Brake Control} (ABC), a simple and deployable explicit congestion control protocol for network paths with time-varying wireless links. ABC routers mark each packet with an ``accelerate'' or ``brake'', which causes senders to slightly increase or decrease their congestion windows. Routers use this feedback to quickly guide senders towards a desired target rate. ABC requires no changes to header formats or user devices, but achieves better performance than XCP. ABC is also incrementally deployable; it operates correctly when the bottleneck is a non-ABC router, and can coexist with non-ABC traffic sharing the same bottleneck link. We evaluate ABC using a Wi-Fi implementation and trace-driven emulation of cellular links. ABC achieves 30-40\% higher throughput than Cubic+Codel for similar delays, and 2.2$\times$ lower delays than BBR on a Wi-Fi path. On cellular network paths, ABC achieves 50\% higher throughput than Cubic+Codel.


\end{abstract}

\section{Introduction}
\label{s:introduction}

This paper proposes a new explicit congestion control protocol for network paths with wireless links. Congestion control on such paths is challenging because of the rapid time variations of the link capacity. Explicit control protocols like XCP~\cite{xcp} and RCP~\cite{rcp} can in theory provide superior performance on such paths compared to end-to-end~\cite{cubic,newreno,bbr,vegas,sprout, verus, copa, vivace} or active queue management (AQM)~\cite{codel, pie} approaches (\S\ref{s:case_explicit}). Unlike these approaches, explicit control protocols enable the wireless router to directly specify a target rate for the sender, signaling both rate decreases and rate increases based on the real-time link capacity.





However, current explicit control protocols have two limitations, one conceptual and the other practical. First, existing explicit protocols were designed for fixed-capacity links; we find that their control algorithms are sub-optimal on time-varying wireless links. Second, they require major changes to packet headers, routers, and endpoints to deploy on the Internet.

Our contribution is a simple and deployable protocol, called Accel-Brake Control (ABC), that overcomes these limitations, building on concepts from a prior position paper~\cite{goyal2017rethinking}.
In \name{} (\S\ref{s:design}), a wireless router marks each packet with one bit of feedback corresponding to either {\em accelerate} or {\em brake} based on a measured estimate of the current link rate. {\color{black}Upon receiving this feedback via an ACK from the receiver, the sender increases its window by one on an accelerate (sends two packets in response to the ACK), and decreases it by one on a brake (does not send any packet).} This simple mechanism allows the router to signal a large dynamic range of window size changes within one RTT: from throttling the window to 0, to doubling the window.

\cut{Upon receiving this feedback via an ACK from the receiver, the sender either increases its window by sending two packets (on accelerate), or decreases it by not sending any packets (on brake). This simple mechanism allows the router to signal a large dynamic range of window size changes within one RTT: from throttling the window to 0, to doubling the window.}

Central to \name's performance is a novel control algorithm that helps routers provide very accurate feedback on time-varying links. Existing explicit schemes like XCP and RCP calculate their feedback by comparing the current {\em enqueue rate} of packets to the link capacity. An \name router, however, compares the {\em dequeue rate} of packets from its queue to the link capacity to mark accelerates or brakes. This change is rooted in the observation that, for an ACK-clocked protocol like \name, the current dequeue rate of packets at the router provides an accurate prediction of the future incoming rate of packets, one RTT in advance. In particular, if the senders maintain the same window sizes in the next RTT, they will send one packet for each ACK, and the incoming rate in one RTT will be equal to the current dequeue rate. Therefore, rather than looking at the current enqueue rate, the router should signal changes based on the anticipated enqueue rate in one RTT to better match the link capacity. The impact of this subtle change is particularly significant on wireless links, since the enqueue and dequeue rates can differ significantly when the link capacity varies.

\name also overcomes the deployability challenges of prior explicit schemes, since it can be implemented on top of the existing explicit congestion notification (ECN)~\cite{ecn-rfc} infrastructure. We present techniques that enable \name to co-exist with non-\name routers, and to share bandwidth fairly with legacy flows traversing a bottleneck \name router (\S\ref{s:coexistence}).

We have implemented \name on a commodity Wi-Fi router running OpenWrt~\cite{openwrt}. Our implementation (\S\ref{s:Wi-Fi}) reveals an important challenge for implementing explicit protocols on wireless links: how to determine the link rate {\color{black} for a user} at a given time?   The task is complicated by the intricacies of the Wi-Fi MAC's batch scheduling and block acknowledgements. We develop a method to estimate the Wi-Fi link rate and demonstrate its accuracy experimentally. For cellular links, the 3GPP standard~\cite{3GPP} shows how to estimate the link rate; our evaluation uses emulation with cellular packet traces. 



We have experimented with ABC in several wireless network settings. Our results are:
\begin{CompactEnumerate}
    \item In Wi-Fi, compared to Cubic+Codel, Vegas, and Copa, \name achieves 30-40\% higher throughput with similar delays. Cubic, PCC Vivace-latency and BBR incur 70\%--6$\times$ higher 95$^{th}$ percentile packet delay with similar throughput.
    \item The results in emulation over 8 cellular traces are summarized below.  Despite relying on single-bit feedback, \name achieves 2$\times$ lower $95^{th}$ percentile packet delay compared to XCP.
    \item \name bottlenecks can coexist with both \name and non-\name bottlenecks. \name flows achieve high utilization and low queuing delays if the bottleneck router is \name, while switching to Cubic when the bottleneck is a non-\name router.
    \item \name competes fairly with both \name and non-\name flows. In scenarios with both \name and non-\name flows, the difference in average throughput of \name and non-\name flows is under 5\%. 
    \begin{center}
\begin{table}
\small
\centering
\begin{tabular}{ c| c| c }
 Scheme & Norm. Utilization & Norm. Delay (95\%)\\ 
 \hline
 \hline
 ABC & 1 (78\%) & 1 (242ms) \\
  \hline
 XCP & {\color{black}0.97} & {\color{black}2.04}\\
  \hline
 Cubic+Codel & {\color{black}0.67} & {\color{red}0.84}  \\  
  \hline
 Copa & {\color{black}0.66} & {\color{red}0.85}\\  
  \hline
 Cubic & {\color{red}1.18} & {\color{black}4.78}\\  
  \hline
 PCC-Vivace & {\color{red}1.12} & {\color{black}4.93}\\  
  \hline
 BBR & {\color{black}0.96} & {\color{black}2.83}\\  
  \hline
 Sprout & {\color{black}0.55} & {\color{black}1.08}\\
  \hline
 Verus & {\color{black}0.72} & {\color{black}2.01}\\
\end{tabular}
\vspace{-6mm}
\end{table}
\end{center}
\end{CompactEnumerate}

\vspace{-5mm}
\section{Motivation}
\label{s:case_explicit}
Link rates in wireless networks can vary rapidly with time; for example, within one second, a wireless link's rate can both double and halve~\cite{sprout}.\footnote{\color{black}We define the link rate for a user as the rate that user can achieve if it keeps the bottleneck router backlogged (see \S\ref{s:linkrate}).} These variations make it difficult for transport protocols to achieve both high throughput and low delay. Here, we motivate the need for explicit congestion control protocols that provide feedback to senders on both rate increases and decreases based on direct knowledge of the wireless link rate. We discuss why these protocols can track wireless link rates more accurately than end-to-end and AQM-based schemes. Finally, we discuss deployment challenges for explicit control protocols, and our design goals for a deployable explicit protocol for wireless links.


\begin{figure*}
    \centering
        \begin{subfigure}[b]{0.24\textwidth}
        \includegraphics[width=\textwidth]{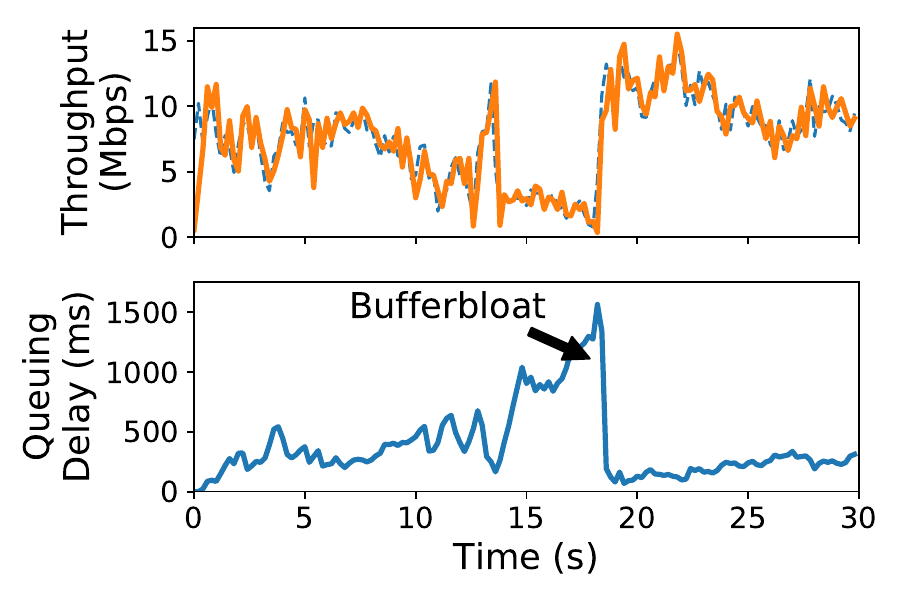}
        \vspace{-6mm}
        \caption{Cubic}
        \label{fig:motivation:cubic}
    \end{subfigure}
    \begin{subfigure}[b]{0.24\textwidth}
        \includegraphics[width=\textwidth]{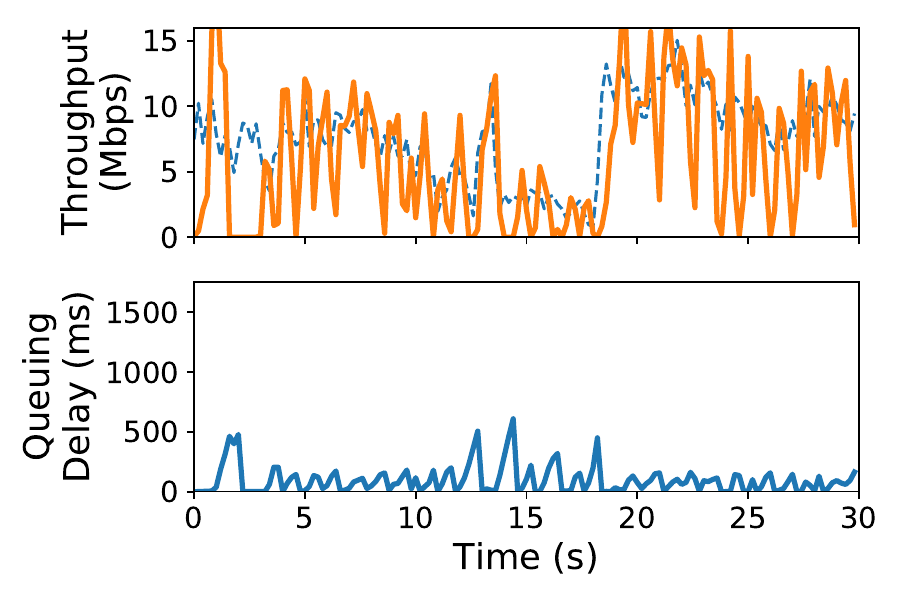}
        \vspace{-6mm}
        \caption{Verus}
        \label{fig:motivation:verus}
    \end{subfigure}
    \begin{subfigure}[b]{0.24\textwidth}
        \includegraphics[width=\textwidth]{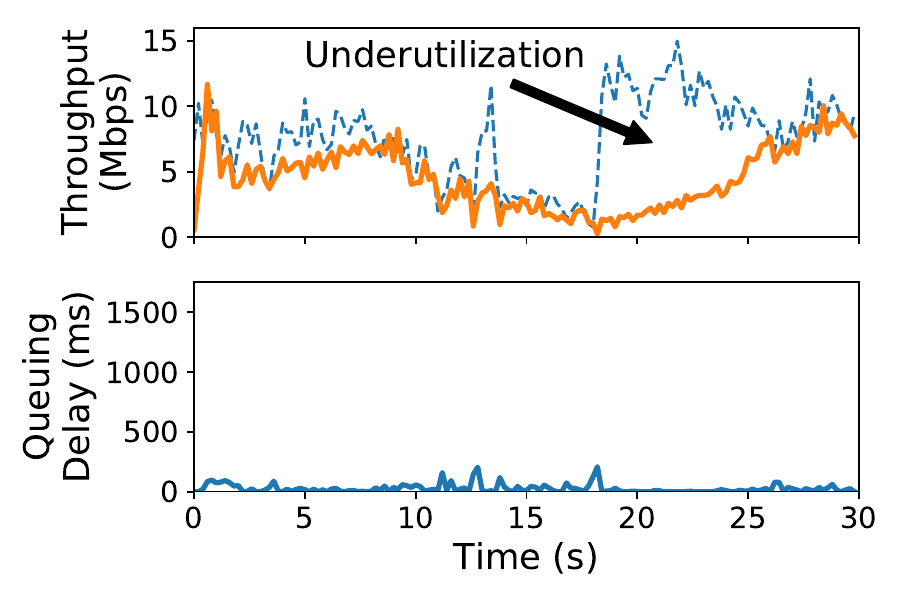}
        \vspace{-6mm}
        \caption{Cubic+CoDel}
        \label{fig:motivation:codel}
    \end{subfigure}
        \begin{subfigure}[b]{0.24\textwidth}
        \includegraphics[width=\textwidth]{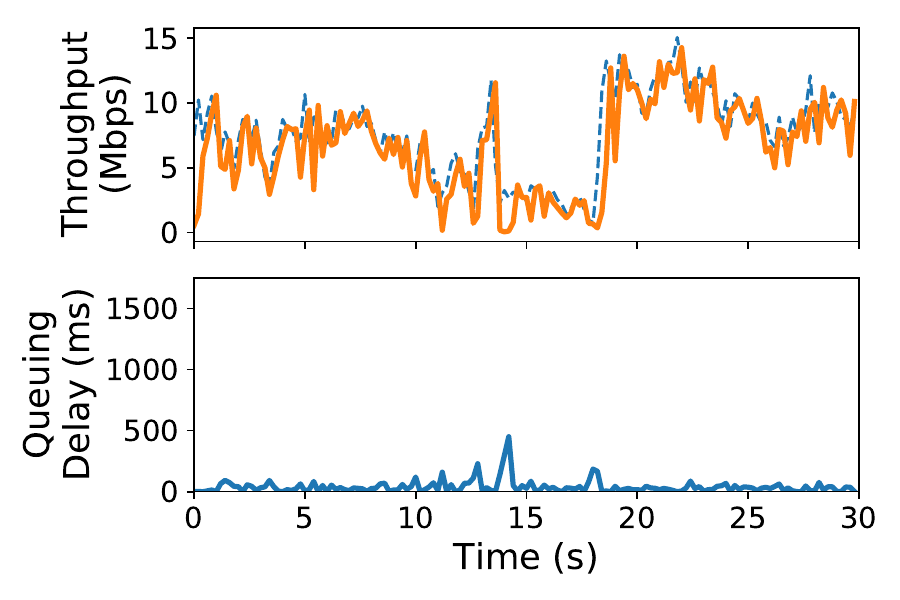}
        \vspace{-6mm}
        \caption{\name}
        \label{fig:motivation:abc}
    \end{subfigure}
    \vspace{-3mm}
    \caption{\small  {\bf Performance on a emulated cellular trace ---} The dashed blue in the top graph represents link capacity, the solid orange line represents the achieved throughput. Cubic has high utilization but has very high delays (up to 1500 milliseconds). Verus has large rate variations and incurs high delays. Cubic+CoDel reduces queuing delays significantly, but leaves the link underutilized when capacity increases. \name achieves close to 100\% utilization while maintaining low queuing delays (similar to that of Cubic+CoDel).}
    \label{fig:motivation}
    \vspace{-6mm}
\end{figure*}

\smallskip
\noindent\textbf{Limitations of end-to-end congestion control:} Traditional end-to-end congestion control schemes like Cubic~\cite{cubic} and NewReno~\cite{newreno} rely on packet drops to infer congestion and adjust their rates. Such schemes tend to fill up the buffer, causing large queuing delays, especially in cellular networks that use deep buffers to avoid packet loss~\cite{sprout}. \Fig{motivation:cubic} shows performance of Cubic on an LTE link, emulated using a LTE trace with Mahimahi~\cite{mahimahi}. The network round-trip time is 100~ms and the buffer size is set to 250 packets. Cubic causes significant queuing delay, particularly when the link capacity drops.


Recent proposals such as BBR~\cite{bbr}, PCC-Vivace~\cite{vivace} and Copa~\cite{copa} use RTT and send/receive rate measurements to estimate the available link rate more accurately. Although these schemes are an improvement over loss-based schemes, their performance is far from optimal on highly-variable links. Our experiments show that they either cause excessive queuing or underutilize the link capacity (e.g., see \Fig{througput-delay}). Sprout~\cite{sprout} and Verus~\cite{verus} are two other recent end-to-end protocols designed specifically for cellular networks. 
They also have difficulty tracking the link rate accurately; depending on parameter settings, they can be too aggressive (causing large queues) or too conservative (hurting utilization). For example, \Fig{motivation:verus} shows how Verus performs on the same LTE trace as above.


The fundamental challenge for any end-to-end scheme is that to estimate the link capacity, it must utilize the link fully and build up a queue. When the queue is empty, signals such as the RTT and send/receive rate do not provide information about the available capacity. Therefore, in such periods, all end-to-end schemes must resort to some form of ``blind'' rate increase. But for networks with a large dynamic range of rates, it is very difficult to tune this rate increase correctly: if it is slow, throughput suffers, but making it too fast causes overshoots and large queuing delays.\footnote{BBR attempts to mitigate this problem by periodically increasing its rate in short pulses, but our experiments show that BBR frequently overshoots the link capacity with variable-bandwidth links, causing excessive queuing {\color{black}(see \App{comp_bbr})}.}  For schemes that attempt to limit queue buildup, periods in which queues go empty (and a blind rate increase is necessary) are common; they occur, for example, following a sharp increase in link capacity. 
 


\smallskip
\noindent{\bf AQM schemes do not signal increases:} AQM schemes like RED~\cite{red}, PIE~\cite{pie} and CoDel~\cite{sfqCoDel} can be used to signal congestion (via ECN or drops) before the buffer fills up at the bottleneck link, reducing delays. However, AQM schemes do not signal rate increases. When capacity increases, the sender must again resort to a blind rate increase. \Fig{motivation:codel} shows how CoDel performs when the sender is using Cubic. Cubic+CoDel reduces delays by 1 to 2 orders of magnitude compared to Cubic alone but leaves the link underutilized when capacity increases.


Thus, we conclude that, both end-to-end and AQM-based schemes will find it difficult to track time-varying wireless link rates accurately. Explicit control schemes, such as XCP~\cite{xcp} and RCP~\cite{rcp} provide a compelling alternative. \cut{In these schemes, }The router provides multiple bits of feedback per packet to senders based on direct knowledge of the wireless link capacity. By telling senders precisely how to increase or decrease their rates, explicit schemes can quickly adapt to time-varying links, in principle, within an RTT of link capacity changes. 


\smallskip
\noindent{\bf Deployment challenges for explicit congestion control:} Schemes like XCP and RCP require major changes to packet headers, routers, and endpoints. Although the changes are technically feasible, in practice, they create significant deployment challenges. For instance, these protocols require new packet fields to carry multi-bit feedback information. IP or TCP options could in principle be used for these fields. But many wide-area routers drop packets with IP options~\cite{fonseca2005ip}, and using TCP options creates problems due to middleboxes~\cite{honda2011still} and IPSec encryption~\cite{seo2005security}. Another important challenge is co-existence with legacy routers and legacy transport protocols. To be deployable, an explicit protocol must handle scenarios where the bottleneck is at a legacy router, or when it shares the link with standard end-to-end protocols like Cubic.


\smallskip
\noindent{\bf Design goals:} In designing \name, we targeted the following properties:
\begin{CompactEnumerate}
    \item {\em Control algorithm for fast-varying wireless links:} Prior explicit control algorithms like XCP and RCP were designed for fixed-capacity links. We design \name's control algorithm specifically to handle  the rapid bandwidth variations and packet transmission behavior of wireless links (e.g., frame batching at the MAC layer). 
    
    \item {\em No modifications to packet headers:} \name repurposes the existing ECN~\cite{ecn-rfc} bits to signal both increases and decreases to the sender's congestion window. By spreading feedback over a sequence of 1-bit signals per packet, ABC routers precisely control sender congestion windows over a large dynamic range. 
    

    \item {\em Coexistence with legacy bottleneck routers:} \name is robust to scenarios where the bottleneck link is not the wireless link but a non-ABC link elsewhere on the path. Whenever a non-\name router becomes the bottleneck, \name senders ignore window increase feedback from the wireless link, and ensure that they send no faster than their fair share of the bottleneck link.
    
    \item {\em Coexistence with legacy transport protocols:} \name routers ensure that ABC and non-ABC flows share a wireless bottleneck link fairly. To this end, ABC routers separate ABC and non-ABC flows into two queues, and use a simple algorithm to schedule packets from these queues. ABC makes no assumptions about the congestion control algorithm of non-ABC flows, is robust to the presence of short or application-limited flows, and requires  a small amount of state at the router.


\end{CompactEnumerate}

\Fig{motivation:abc} shows \name on the same emulated LTE link. Using only one bit of feedback per packet, the \name flow is able to track the variations in bottleneck link closely, achieving both high throughput and low queuing delay.

\section{Design}
\label{s:design}

\name is a window-based protocol: the sender limits the number of packets in flight to the current congestion window. Window-based protocols react faster to the sudden onset of congestion than rate-based schemes~\cite{bansal2001}. On a wireless link, when the capacity drops and the sender stops receiving ACKs, \name will stop sending packets immediately, avoiding further queue buildup. In contrast, a rate-based protocol would take time to reduce its rate and may queue up a large number of packets at the bottleneck link in the meantime.

\name senders adjust their window size based on explicit feedback from \name routers. An \name router uses its current estimate of the link rate and the queuing delay to compute a {\em target rate}. The router then sets one bit of feedback in each packet to guide the senders towards the target rate. Each bit is echoed to a sender by a receiver in an ACK, and it signals either a one-packet increase (``accelerate'') or a one-packet decrease (``brake'') to the sender's congestion window.

\subsection{The ABC Protocol}
We now present \name's design starting with the case where all routers are \name-capable and all flows use \name. We later discuss how to extend the design to handle non-ABC routers and scenarios with competing non-ABC flows.

\subsubsection{ABC Sender}
\label{s:abc-sender}

On receiving an ``accelerate'' ACK, an \name sender increases its congestion window by 1 packet. This increase results in two packets being sent, one in response to the ACK and one due to the window increase. On receiving a ``brake,'' the sender reduces its congestion window by 1 packet, preventing the sender from transmitting a new packet in response to the received ACK. As we discuss in \S\ref{s:fairness}, the sender also performs an additive increase of 1 packet per RTT to achieve fairness. For ease of exposition, let us ignore this additive increase for now.

Though each bit of feedback translates to only a small change in the congestion window, when aggregated over an RTT, the feedback can express a large dynamic range of window size adjustments. For example, suppose a sender's window size is $w$, and the router marks accelerates on a fraction $f$ of packets in that window. Over the next RTT, the sender will receive $w \cdot f$ accelerates and $w - w \cdot f$ brakes. Then, the sender's window size one RTT later will be $w + w f - (w - w f) = 2wf$ packets. Thus, in one RTT, an ABC router can vary the sender's window size between zero ($f$ = 0) and double its current value ($f = 1$). The set of achievable window changes for the next RTT depends on the number of packets in the current window $w$; the larger $w$, the higher the granularity of control.

In practice, \name senders increase or decrease their congestion window by the number of newly acknowledged bytes covered by each ACK. Byte-based congestion window modification is a standard technique in many TCP implementations~\cite{allman2003tcp}, and it makes \name robust to variable packet sizes and delayed, lost, and partial ACKs. For simplicity, we describe the design with packet-based window modifications in this paper.

\subsubsection{\name Router}
\label{s:abc-router}
\noindent{\bf Calculating the target rate:} 
\name routers compute the target rate $tr(t)$ using the following rule:
\vspace{-2.5mm}
\begin{equation}
tr(t) = \eta {\mu(t)} - \frac{\mu(t)}{\delta} (x(t) - d_{t})^{+},
\label{eq:abctargetrule}
\vspace{-3mm}
\end{equation}
where $\mu(t)$ is the link capacity, $x(t)$ is the observed queuing delay, $d_{t}$ is a pre-configured delay threshold, $\eta$ is a constant less than 1,  $\delta$ is a positive constant (in units of time), and $y^{+}$ is $\max(y,0)$. This rule has the following interpretation. When queuing delay is low ($x(t) < d_t$), \name sets the target rate to $\eta \mu(t)$, for a value of $\eta$ slightly less than 1 (e.g., $\eta = 0.95$). By setting the target rate a little lower than the link capacity, \name aims to trade a small amount of bandwidth for large reductions in delay,  similar to prior work~\cite{jain1996congestion,hull,avq}. However, queues can still develop due to rapidly changing link capacity and the 1 RTT of delay it takes for senders to achieve the target rate. \name uses the second term in Equation~\eqref{eq:abctargetrule} to drain queues. Whenever $x(t) > d_t$, this term reduces the target rate by an amount that causes the queuing delay to decrease to $d_t$ in at most $\delta$ seconds. 

The threshold $d_t$ ensures that the target rate does not react to small increases in queuing delay. This is important because wireless links often schedule packets in batches. Queuing delay caused by batch packet scheduling does not imply congestion, even though it occurs persistently. To prevent target rate reductions due to this delay, $d_{t}$ must be configured to be greater than the average inter-scheduling time at the router.

\name's target rate calculation requires an estimate of the underlying link capacity, $\mu(t)$. In \Sec{linkrate}, we discuss how to estimate the link capacity in cellular and WiFi networks, and we present an implementation for WiFi.

\begin{figure}[t]
    \centering
    \begin{subfigure}[t]{0.22\textwidth}
        \includegraphics[width=\textwidth]{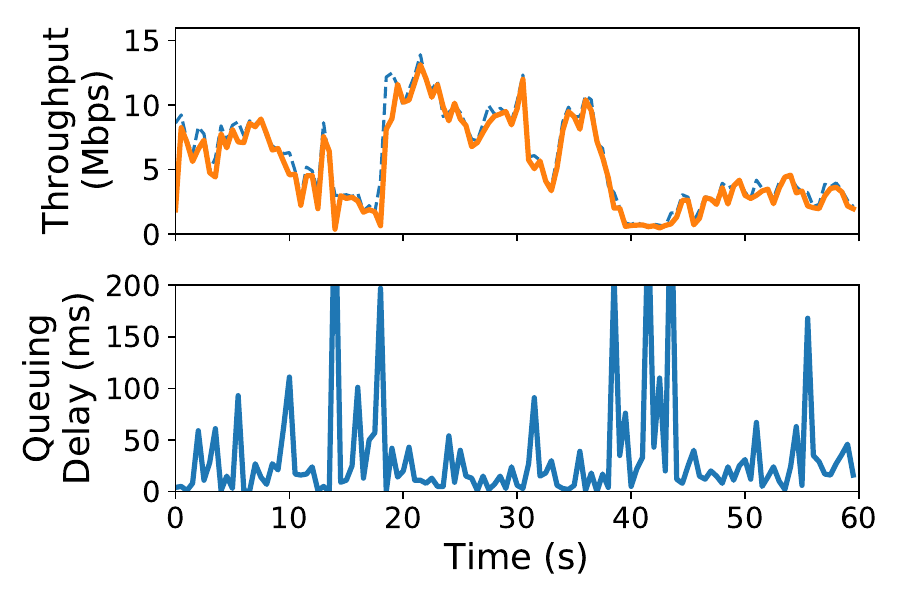}
        \vspace{-6.75mm}
        \caption{Dequeue}
        \label{fig:illustration:dq}
    \end{subfigure}
    \begin{subfigure}[t]{0.22\textwidth}
        \includegraphics[width=\textwidth]{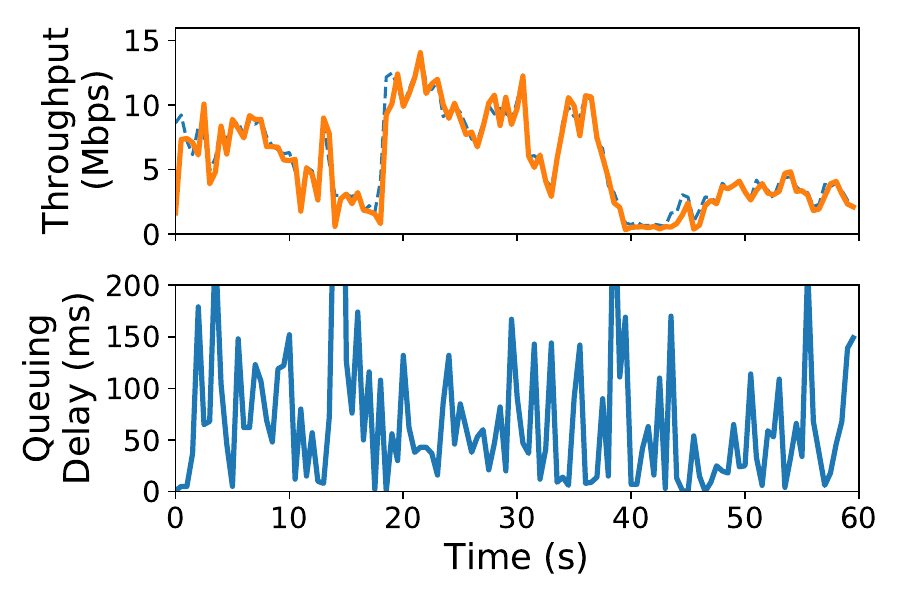}
        \vspace{-6.75mm}
        \caption{Enqueue}
        \label{fig:illustration:eq}
    \end{subfigure}
    \vspace{-3.25mm}
    \caption{\small{\bf Feedback ---} Calculating $f(t)$ based on enqueue rate increases $95^{th}$ percentile queuing delay by 2$\times$.}
    \label{fig:illustration}
    \vspace{-6mm}
\end{figure}

\smallskip
\noindent {\bf Packet marking:} To achieve a target rate, $tr(t)$, the router computes the fraction of packets, $f(t)$, that should be marked as accelerate. Assume that the current {\em dequeue rate}\,---\,the rate at which the router transmits packets\,---\,is $cr(t)$. If the accelerate fraction is $f(t)$, for each packet that is ACKed, the sender transmits $2f(t)$ packets on average. Therefore, after 1 RTT, the {\em enqueue rate}\,---\,the rate at which packets arrive to the router\,---\,will be $2cr(t)f(t)$. To achieve the target rate, $f(t)$ must be chosen such that $2cr(t)f(t)$ is equal to $tr(t)$. Thus, $f(t)$ is given by:
\vspace{-3mm}
\begin{equation}
\vspace{-2mm} f(t) = \min\Big\{\frac{1}{2} \cdot \frac{tr(t)}{cr(t)}, 1 \Big\}.
\label{eq:f}
\end{equation}

An important consequence of the above calculation is that $f(t)$ is computed based on the {\em dequeue} rate. Most explicit protocols compare the enqueue rate to the link capacity to determine the feedback (e.g., see XCP~\cite{xcp}). 

\name uses the dequeue rate instead to exploit the ACK-clocking property of its window-based protocol. Specifically, Equation~\eqref{eq:f} accounts for the fact that when the link capacity changes (and hence the dequeue rate changes), the rate at the senders changes automatically within 1 RTT because of ACK clocking. \Fig{illustration} demonstrates that computing $f(t)$ based on the dequeue rate at the router enables \name to track the link capacity much more accurately than using the enqueue rate. 

\name recomputes $f(t)$ on every dequeued packet, using measurements of $cr(t)$ and $\mu(t)$ over a sliding time window of length $T$. Updating the feedback on every packet allows \name to react to link capacity changes more quickly than schemes that use periodic feedback updates (e.g., XCP and RCP).


Packet marking can be done deterministically or probabilistically. To limit burstiness, \name uses the deterministic method in Algorithm~\ref{alg:abc}. The variable {\tt token} implements a token bucket that is incremented by $f(t)$ on each outgoing packet (up to a maximum value {\tt tokenLimit}), and decremented when a packet is marked accelerate. To mark a packet accelerate, {\tt token} must exceed 1. This simple method ensures that no more than a fraction $f(t)$ of the packets are marked accelerate. 



\vspace{-2mm}
\begin{algorithm}

\small
\SetAlgoLined
\caption{Packet marking at an \name{} router.\vspace{-13mm}}
\hrule
token $=$ 0\;
\For{each outgoing packet}{
    calculate $f(t)$ using Equation~\eqref{eq:f}\;
    token = min(token $+ f(t)$, tokenLimit)\;
    \If{packet marked with accelerate}{
        \eIf{token $>$ 1}{
            token = token $-$ 1\;
            mark accelerate\;
        }{
            mark brake\;
        }
    }
}
\label{alg:abc}
\end{algorithm}

\vspace{-4mm}
\smallskip
\noindent{\bf Multiple bottlenecks:} An \name flow may encounter multiple ABC routers on its path. An example of such a scenario is when two smartphone users communicate over an \name-compliant cellular network. Traffic sent from one user to the other will traverse a cellular uplink and cellular downlink, both of which could be the bottleneck. To support such situations, an \name sender should send traffic at the smallest of the router-computed target rates along their path.  To achieve this goal, each packet is initially marked accelerate by the sender. \name routers may change a packet marked accelerate to a brake, but not vice versa (see Algorithm~\ref{alg:abc}). This rule guarantees that an \name router can unilaterally reduce the fraction of packets marked accelerate to ensure that its target rate is not exceeded, but it cannot increase this fraction. Hence the fraction of packets marked accelerate will equal the minimum $f(t)$ along the path. 

\subsubsection{Fairness}
\label{s:fairness_description}
\begin{figure}
    \centering
    \begin{subfigure}[t]{0.23\textwidth}
        \includegraphics[width=\textwidth]{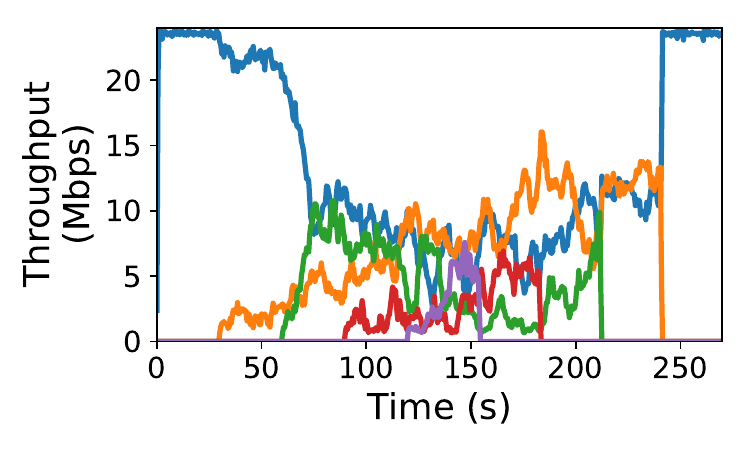}
        \vspace{-6mm}
        \caption{\name w/o AI}
        \label{fig:fairness:no_ai}
    \end{subfigure}
    \begin{subfigure}[t]{0.23\textwidth}
        \includegraphics[width=\textwidth]{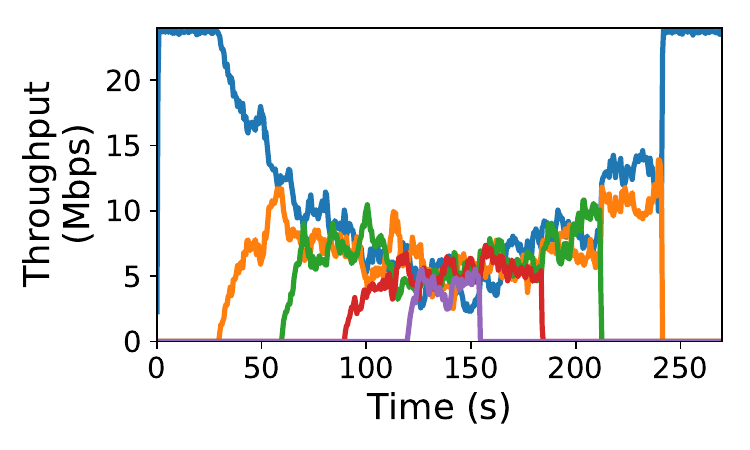}
        \vspace{-6mm}
        \caption{\name with AI}
        \label{fig:fairness:ai}
    \end{subfigure}
    \vspace{-3.25mm}
    \caption{\small {\bf Fairness among competing \name flows ---} 5 flows with the same RTT start and depart one-by-one on a 24~Mbit/s link. The additive-increase (AI) component leads to fairness.}
    \label{fig:fairness}
    \vspace{-5mm}
\end{figure}

Multiple \name flows sharing the same bottleneck link should be able to compete fairly with one another. However, the basic window update rule described in ~\Sec{abc-sender} is a multiplicative-increase/multiplicative-decrease (MIMD) strategy,\footnote{All the competing \name senders will observe the same accelerate fraction, $f$, on average. Therefore, each flow will update its congestion window, $w$, in a multiplicative manner, to $2fw$, in the next RTT.} which does not provide fairness among contending flows (see \Fig{fairness:no_ai} for an illustration). To achieve fairness, we add an additive-increase (AI) component to the basic window update rule. Specifically,  \name senders adjust their congestion window on each ACK as follows:

\vspace{-2.5mm}
\begin{equation}
w \leftarrow 
\begin{cases}
      w + 1 + 1/w & \text{if accelerate} \\
      w - 1 + 1/w & \text{if brake}
    \end{cases}
    \label{eq:abc_cwnd_rule}
\vspace{-1mm}
\end{equation}

This rule increases the congestion window by 1 packet each RTT, in addition to reacting to received accelerate and brake ACKs. This additive increase, coupled with \name's MIMD response, makes \name a multiplicative-and-additive-increase/multiplicative-decrease (MAIMD) scheme. Chiu and Jain~\cite{chiujain} proved that MAIMD schemes converge to fairness (see also~\cite{akella2002exploring}). \Fig{fairness:ai} shows how with an AI component, competing \name flows achieve fairness.

To give intuition, we provide a simple informal argument for why including additive increase gives \name fairness. Consider $N$ \name flows sharing a link, and suppose that in steady state, the router marks a fraction $f$ of the packets accelerate, and the window size of flow $i$ is $w_i$. To be in steady state, each flow must send 1 packet on average for each ACK that it receives. Now consider flow $i$. It will send $2f+1/w_i$ packets on average for each ACK: $2f$ for the two packets it sends on an accelerate (with probability $f$), and $1/w_i$ for the extra packet it sends every $w_i$ ACKs. Therefore, to be in steady state, we must have: $2f + 1/w_i = 1 \implies w_i = 1 / (1-2f)$. This shows that the steady-state window size for all flows must be the same, since they all observe the same fraction $f$ of accelerates. Hence, with equal RTTs, the flows will have the same throughput, and otherwise their throughput will be inversely proportional to their RTT. Note that the RTT unfairness in ABC is similar to that of schemes like Cubic, for which the throughput of a flow is inversely proportional to its RTT. In ~\Sec{flow_fairness}, we show experiments where flows have different RTTs.

\subsubsection{Stability Analysis}
\label{s:stability_analysis_new}
\name's stability depends on the values of $\eta$ and $\delta$. $\eta$ determines the target link utilization, while $\delta$ controls how long it will take for a queue\cut{ buildup } to drain. In Appendix~\ref{app:stability}, we prove the following result for a fluid model of the \name control loop.
\begin{theorem}
\vspace{-1.25mm}
Consider a single \name link, traversed by $N$ \name flows. Let $\tau$ be the maximum round-trip propagation delay of the flows. \name is {\em globally asymptotically stable} if
\vspace{-1.4mm}
\begin{align}
    \delta > \frac{2}{3} \cdot \tau.
\vspace{-8mm}
\end{align}
Specifically, if $\mu(t) = \mu$ for $t > t_0$ (i.e., the link capacity stops changing after some time $t_0$), the enqueue/dequeue rate and the queuing delay at the \name router will converge to certain values $r^*$ and $x^*$ that depend on the system parameters and the number of flows. In all cases: $\eta \mu < r^* \leq \mu$.
\label{thm:stablity}
\vspace{-3mm}
\end{theorem}
This stability criterion is simple and intuitive. It states that $\delta$ should not be much smaller than the RTT (i.e, the feedback delay). If $\delta$ is very small, \name reacts too forcefully to queue build up, causing under-utilization and  oscillations.\footnote{Interestingly, if the sources do not perform additive increase or if the additive increase is sufficiently ``gentle,'' \name is stable for any value of $\delta$. See the proof in Appendix~\ref{app:stability} for details.} Increasing $\delta$ well beyond $2/3\tau$ improves the stability margins of the feedback loop, but hurts responsiveness. In our experiments, we used $\delta=133$ ms for a propagation RTT of $100$ ms. 
\section{Coexistence}
\label{s:coexistence}

An \name flow should be robust to presence of non-\name bottlenecks on its path and share resources fairly with non-\name flows sharing the \name router. 


\subsection{Deployment with non-\name{} Routers}
\label{s:multiple-bottleneck}

An \name{} flow can encounter both \name and non-\name routers on its path. For example, a Wi-Fi user's traffic may traverse both a Wi-Fi router (running \name{}) and an ISP router (not running \name{}); either router could be the bottleneck at any given time. \name flows must therefore be able to detect and react to traditional congestion signals---both drops and ECN---and they must determine when to ignore accelerate feedback from \name routers because the bottleneck is at a non-ABC router. 

We augment the \name sender to maintain two congestion windows, one for tracking the available rate on \name routers ($w_{\mbox{abc}}$), and one for tracking the rate on non-\name{} bottlenecks ($w_{\mbox{nonabc}}$). $w_{\mbox{abc}}$ obeys accelerates/brakes using Equation~\eqref{eq:abc_cwnd_rule}, while $w_{\mbox{nonabc}}$ follows a rule such as Cubic~\cite{cubic} and responds to drop and ECN signals.\footnote{We discuss how \name senders distinguish between accelerate/brake and ECN marks in \S\ref{s:deployment}.} An \name sender must send packets to match the lower of the two windows.\footnote{A concurrent work~\cite{kramer2019towards} proposed a similar methodology for this problem.} 
Our implementation mimics Cubic for the non-\name method, but other methods could also be emulated.

With this approach, the window that is not the bottleneck could become large. For example, when a non-\name router is the bottleneck, the \name router will continually send accelerate signals, causing $w_{\mbox{abc}}$ to grow.
If the \name router later becomes the bottleneck, it will temporarily incur large queues. To prevent this problem, \name senders cap both $w_{\mbox{abc}}$ and $w_{\mbox{nonabc}}$ to $2\times$ the number of in-flight packets. 

 \begin{figure}[t]
     \centering
     \includegraphics[width=0.9\columnwidth]{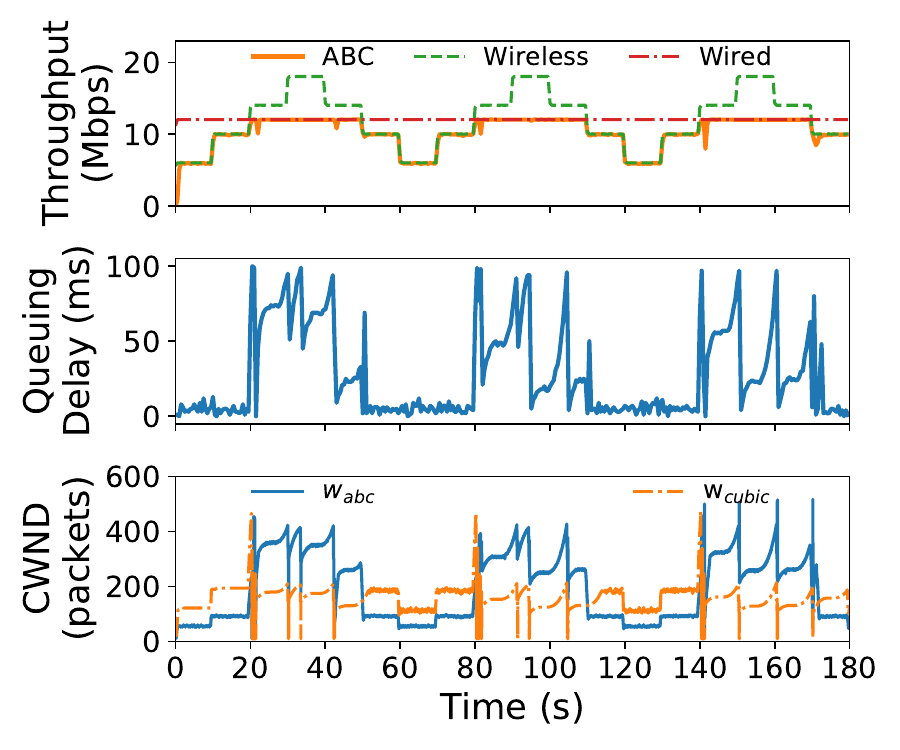}
     \vspace{-4.5mm}
     \caption{\small {\bf Coexistence with non-\name bottlenecks ---} When the wired link is the bottleneck, \name becomes limited by $w_{\mbox{cubic}}$ and behaves like a Cubic flow. When the wireless link is the bottleneck, \name uses $w_{abc}$ to achieve low delays and high utilization.}
     \label{fig:ts_droptail_cell}
     \vspace{-5mm}
\end{figure}

\Fig{ts_droptail_cell} shows the throughput and queuing delay for an \name flow traversing a path with an \name-capable wireless link and a wired link with a droptail queue. For illustration, we vary the rate of the wireless link in a series of steps every 5 seconds. Over the experiment, the bottleneck switches between the wired and wireless links several times. \name is able to adapt its behavior quickly and accurately. Depending on which link is the bottleneck, either $w_{\mbox{nonabc}}$ (i.e., $w_{\mbox{cubic}}$) or $w_{\mbox{abc}}$ becomes smaller and controls the rate of the flow. When the wireless link is the bottleneck, \name maintains low queuing delay, whereas the queuing delay exhibits standard Cubic behavior when the wired link is the bottleneck. 
$w_{\mbox{cubic}}$ does not limit \name's ability to increase its rate when the wireless link is the bottleneck. At these times (e.g., around the 70~s mark), as soon on $w_{\mbox{abc}}$ increases, the number of in-flight packets and the cap on $w_{\mbox{cubic}}$ increases, and $w_{\mbox{cubic}}$ rises immediately.

\subsection{Multiplexing with ECN Bits}
\label{s:deployment}




IP packets have two ECN-related bits: ECT and CE. These two bits are traditionally interpreted as follows:
\begin{center}
\vspace{-2mm}
\begin{tabular}{ c c c }
\small
 ECT & CE & Interpretation  \\ \hline
 0 & 0 & Non-ECN-Capable Transport\\  
 0 & 1 & ECN-Capable Transport ECT(1)\\
 1 & 0 & ECN-Capable Transport ECT(0)\\  
 1 & 1 & ECN set 
 \vspace{-2mm}
\end{tabular}
\end{center}
Routers interpret both 01 and 10 to indicate that a flow is ECN-capable, and routers change those bits to 11 to mark a packet with ECN. Upon receiving an ECN mark (11), the receiver sets the {\em ECN Echo (ECE)} flag to signal congestion to the sender. \name reinterprets the ECT and CE bits as follows:


\begin{center}
\vspace{-2mm}
\begin{tabular}{ c c c }
\small
 ECT & CE & Interpretation  \\ \hline
 0 & 0 & Non-ECN-Capable Transport\\  
 0 & 1 & \textbf{Accelerate}\\
 1 & 0 & \textbf{Brake}\\  
 1 & 1 & ECN set
 \vspace{-2mm}
\end{tabular}
\end{center}
\name send all packets with accelerate (01) set, and \name routers signal brakes by flipping the bits to 10. Both 01 and 10 indicate an ECN-capable transport to ECN-capable legacy routers, which will continue to use (11) to signal congestion.

With this design, receivers must be able to echo both standard ECN signals and accelerates/brakes for \name. Traditional ECN feedback is signaled using the ECE flag. For \name feedback, we repurpose the NS (nonce sum) bit, which was originally proposed to ensure ECN feedback integrity~\cite{nonce} but has been reclassified as historic~\cite{accurate-ecn} due to lack of deployment. Thus, it appears possible to deploy \name with only simple modifications to TCP receivers.

\noindent {\bf Deployment in proxied networks:} Cellular networks commonly split TCP connections and deploy proxies at the edge~\cite{wang2011untold,ravindranath2013timecard}. Here, it is unlikely that any non-\name router will be the bottleneck and interfere with the accel-brake markings from the \name router. In this case, deploying \name may not require any modifications to today's TCP ECN receiver. \name senders (running on the proxy) can  use either 10 or 01 to signal an accelerate, and routers can use 11 to indicate a brake.  The TCP receiver can echo this feedback using the ECE flag. 

\subsection{Non-\name flows at an \name{} Router}
\label{s:fairness}
\name flows are potentially at a disadvantage when they share an \name bottleneck link with non-\name flows.\footnote{\name and non-\name flows may also share a non-\name  link, but in such cases, \name flows will behave like Cubic and compete fairly with other traffic.} If the non-\name flows fill up queues and increase queuing delay, the \name router will reduce \name's target rate. To ensure fairness in such scenarios, \name routers isolate \name and non-\name packets in separate queues. 

We assume that \name routers can determine whether a packet belongs to an \name flow. In some deployment scenarios, this is relatively straightforward. For example, in a cellular network deployment with TCP proxies at the edge of the network~\cite{wang2011untold,ravindranath2013timecard}, the operator can deploy \name at the proxy, and configure the base station to assume that all traffic from the proxy's IP address uses \name. Other deployment scenarios may require \name senders to set a predefined value in a packet field like the IPv6 flow label or the IPv4 IPID. 

The \name router assigns weights to the \name and non-\name queues, respectively, and it schedules packets from the queues in proportion to their weights. In addition, \name's target rate calculation considers only \name's share of the link capacity (which is governed by the weights). The challenge is  to set the weights to ensure that the average throughput of long-running \name and non-\name flows is the same, no matter how many flows there are.

Prior explicit control schemes address this problem using the TCP loss-rate equation (XCP) or by estimating the number of flows with Zombie Lists (RCP). Relying on the TCP equation requires a sufficient loss rate and does not handle flows like BBR. RCP's approach does not handle short flows. When one queue has a large number of short flows (and hence a low average throughput), RCP increases the weight of that queue. However, the short flows cannot send faster, so the extra bandwidth is taken by long-running flows in the same queue, which get more throughput than long-running flows in the other queue (see \S\ref{ss:flow_fairness} for experimental results).

To overcome these drawbacks, 
a \name{} router measures the average rate of the $K$ largest flows in each queue using the Space Saving Algorithm~\cite{metwally2005efficient}, which requires $\mathcal{O}(K)$ space. It considers any remaining flow in either queue to be short, and it calculates the total rate of the short flows in each queue by subtracting the rate of the largest $K$ flows from the queue's aggregate throughput. \name uses these rate measurements to estimate the rate demands of the flows. Using these demands, \name periodically computes the max-min fair rate allocation for the flows, and it sets the weight of each of the two queues to be equal to the total max-min rate allocation of its component flows. This algorithm ensures that long-running flows in the two queues achieve the same average rate, while accounting for demand-limited short flows. 

To estimate the demand of the flows, the \name router assumes that the demand for the top $K$ flows in each queue is X\% higher than the current throughput of the flow, and the aggregate demand for the short flows is the same as their throughput. If a top-$K$ flow is unable to increase its sending rate by X\%, its queue's weight will be larger than needed, but any unfairness in weight assignment is bounded by X\%. Small values of $X$ limit unfairness but can slow down convergence to fairness; our experiments use $X=10\%$. 

\section{Estimating Link Rate}
\label{s:linkrate}
We describe how \name routers can estimate the link capacity for computing the target rate (\S\ref{s:abc-router}). We present a technique for Wi-Fi that leverages the inner workings of the Wi-Fi MAC layer, and we discuss options for cellular networks. 

\subsection{Wi-Fi}
\label{s:Wi-Fi}
We describe how an 802.11n access point (AP) can estimate the average link rate. {\color{black} For simplicity, we first describe our solution when there is a single user (client) connected to the AP. Next, we describe the multi-user case.}

{\color{black} We define {\em link rate} as the potential throughput of the user (i.e., the MAC address of the Wi-Fi client) if it was backlogged at the AP, \ie if the user never ran out of packets at the AP. In case the router queue goes empty at the AP, the achieved throughput will be less than the link rate.}

\smallskip
\noindent{\bf Challenges:} A strawman would be to estimate the link rate using the physical layer bit rate selected for each
 transmission, which would depend on the modulation and channel code used for the transmission. Unfortunately, this method will overestimate the link rate as the packet transmission times are governed not only by {\color{black}the bitrate}, but also by delays for additional tasks (e.g., channel contention and retransmissions~\cite{samplerate}). An alternative approach would be to use the fraction of time that the router queue was backlogged as a proxy for link utilization. However, the Wi-Fi MAC's packet batching confounds this approach. Wi-Fi routers transmit packets (frames) in batches; a new batch is transmitted only after receiving an ACK for the last batch. The AP may accumulate packets while waiting for a link-layer ACK; this queue buildup does not necessarily imply that the link is fully utilized. Thus, accurately measuring {\color{black}the link rate} requires a detailed consideration of Wi-Fi's packet transmission protocols. 

\smallskip
\noindent{\bf Understanding batching:} In 802.11n, data frames, also known as MAC Protocol Data Units (MPDUs), are transmitted in batches called A-MPDUs (Aggregated MPDUs). The maximum number of frames that can be included in a single batch, $M$, is negotiated by \cut{each}{\color{black} the} receiver and the router. When \cut{a given}{\color{black} the} user is not backlogged, the router might not have enough data to send a full-sized batch of $M$ frames, but will instead use a smaller batch of size $b < M$. Upon receiving a batch, the receiver responds with a single Block ACK. Thus, at a time $t$, given a batch size of $b$ frames, a frame size of $S$ bits,\footnote{For simplicity, we assume that all frames are of the same size, though our formulas can be generalized easily for varying frame sizes.} and an ACK inter-arrival time (i.e., the time between receptions of consecutive block ACKs) of $T_{IA}(b,t)$, the current dequeue rate, $cr(t)$, may be estimated as
\vspace{-2mm}
\begin{equation}
     cr(t) = \frac{b \cdot S}{T_{IA}(b,t)}.
     \label{eq:inst_throughput}
\end{equation}
\vspace{-4mm}

When the user is backlogged and $b=M$, then $cr(t)$ above will be equal to the link capacity. However, if the user is not backlogged and $b < M$, how can the AP estimate the link capacity? Our approach calculates $\hat{T}_{IA}(M,t)$, the estimated ACK inter-arrival time {\em if the user was backlogged and had sent $M$ frames in the last batch.}

We estimate the link capacity, $\hat{\mu}(t)$, as
\vspace{-2mm}
  \begin{align}
    \hat{\mu}(t) = \frac{M \cdot S}{\hat{T}_{IA}(M,t)}. 
         \label{eq:capacity_backlog}
  \end{align}
\vspace{-4mm}

 \begin{figure}[t]
     \centering
     \includegraphics[width=0.9\columnwidth]{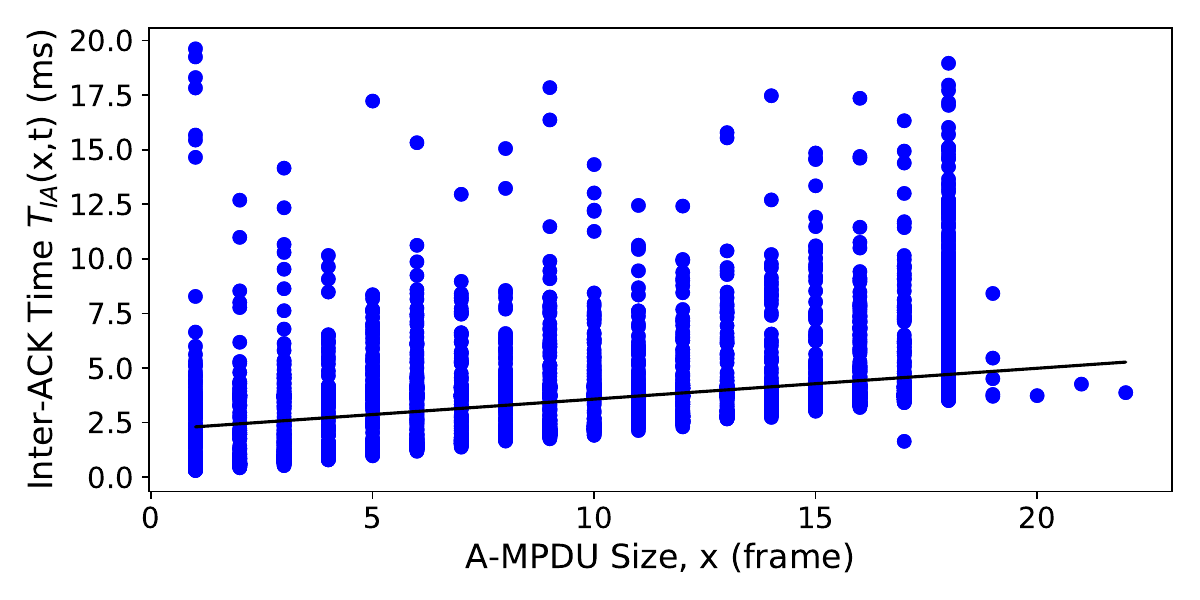}
     \vspace{-4mm}
     \caption{\small {\bf Inter-ACK time v. batch (A-MPDU) size ---} Inter-ACK times for a given batch size exhibits variation. The solid black line represents the average Inter-ACK time. The slope of the line is $S/R$, where $S$ is the frame size in bits and $R$ is the link rate in bits per second.}
     \label{fig:ampdu_burst}
     \vspace{-3mm}
\end{figure}

To accurately estimate $\hat{T}_{IA}(M,t)$, we turn to the relationship between the batch size and ACK inter-arrival time. We can decompose the ACK interval time into the batch transmission time and ``overhead'' time, the latter including physically receiving an ACK, contending for the shared channel, and transmitting the physical layer preamble~\cite{ginzburg2007performance}. Each of these overhead components is independent of the batch size. {\color{black} We denote the overhead time by $h(t)$}. If $R$ is the bitrate used for transmission, the router's ACK inter-arrival time is
\vspace{-1mm}
 \begin{align}
     T_{IA}(b,t) \;&=\; \frac{b \cdot S}{R} + h(t).
     \label{eq:inter_ack}
 \end{align}
 \vspace{-4mm}

\Fig{ampdu_burst} illustrates this relationship empirically. There are two key properties to note. First, for a given batch size, the ACK inter-arrival times vary due to overhead tasks. Second, because the overhead time and batch size are independent, connecting the average values of ACK inter-arrival times across all considered batch sizes will produce a line with slope $S/R$. Using this property {\color{black} along with Equation~\eqref{eq:inter_ack}}, we can estimate the ACK inter-arrival time for a backlogged user as

\vspace{-6mm}
 \begin{align}
    \hat{T_{IA}}(M,t) \;&=\;  \frac{M \cdot S}{R} + h(t)\nonumber\\
     \;&=\; T_{IA}(b,t) + \frac{(M - b) \cdot S}{R} \cdot
    \label{eq:inter-ack-time}
 \end{align}
 \vspace{-6mm}

We can then use $\hat{T}_{IA}(M,t)$ to estimate the link capacity with Equation~\eqref{eq:capacity_backlog}. This computation is performed for each batch transmission when the batch ACK arrives, and passed through a weighted moving average filter over a sliding window of time $T$ to estimate the smoothed time-varying link rate. $T$ must be greater than the inter-ACK time (up to 20 ms in \Fig{ampdu_burst}); we use $T = 40$ ms. Because \name cannot exceed a rate-doubling per RTT, we cap the predicted link rate to double the current rate (dashed slanted line in \Fig{ampdu_burst}). 

 \begin{figure}[t]
     \centering
     \includegraphics[width=0.9\columnwidth]{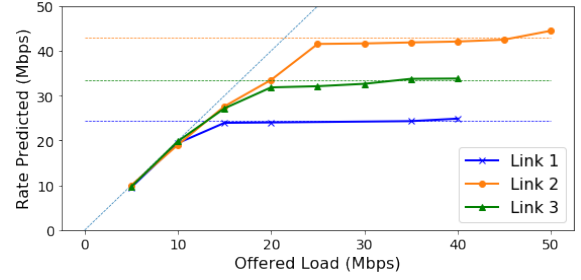}
     \vspace{-3mm}
     \caption{\small {\bf Wi-Fi Link Rate Prediction ---} \name router link rate predictions for a user that was not backlogged and sent traffic at multiple different rates over three different Wi-Fi links. Horizontal lines represent the true link capacity, solid lines summarize the \name router's link capacity prediction (each point is an average over 30 seconds of predictions), and the dashed slanted line represents the prediction rate caps. \name's link rate predictions are within 5\% of the ground truth across {\color{black} most sending rates (given the prediction cap).}}
     \label{fig:prediction}
     \vspace{-5mm}
\end{figure}

 To evaluate the accuracy of our link rate estimates, we transmit data to a single client through our modified \name router (\Sec{prototype implementation}) at multiple different rates over three Wi-Fi links (with different modulation and coding schemes). \Fig{prediction} summarizes the accuracy of the \name router's link rate estimates. With this method, the \name Wi-Fi router is able to predict link rates within 5\% of the true link capacities.

\smallskip
\noindent
\textbf{Extension to multiple users.} {\color{black} In multi-user scenarios, each receiver will negotiate its own maximum batch size ($M$) with the router, and different users can have different transmission rates. We now present two variants of our technique for (1) when the router uses per-user queues to schedule packets of different users, and (2) when the users share a single FIFO (first-in first-out) queue at the router.}

\noindent
\textit{Per-user queues.} {\color{black} In this case each user calculates a separate link rate estimate. Recall that the link rate for a given user is defined as the potential throughput of the user if it was backlogged at the router. To determine the link rate for a user~$x$, we repeat the single-user method for the packets and queue of user~$x$ alone, treating transmissions from other users as overhead time. Specifically,
user $x$ uses Equations~\eqref{eq:inter-ack-time} and~\eqref{eq:capacity_backlog} to compute its link rate ($\hat{\mu}_x(t)$) based on its own values of the bit rate ($R_x$) and maximum batch size ($M_x$). It also computes its current dequeue rate ($cr_{x}(t)$) using Equation~\eqref{eq:inst_throughput} to calculate the accel-brake feedback. The inter-ACK time ($T_{IA_{x}}(b,t)$), is defined as the time between the reception of consecutive block-ACKs for user $x$. Thus, the overhead time ($h_{x}(t)$) includes the time when other users at the same AP are scheduled to send packets. Fairness among different users is ensured via scheduling users out of separate queues.}

\noindent
\textit{Single queue.} {\color{black} In this case the router calculates a single aggregate link rate estimate. The inter-ACK time here is the time between two consecutive block-ACKs, regardless of the user to which the block-ACKs belong to. The router tries to match the aggregate rate of the senders to the aggregate link rate, and uses the aggregate current dequeue rate to calculate accel-brake feedback.}


\subsection{Cellular Networks}
\label{cellular_estimation}

Cellular networks schedule users from separate queues to ensure inter-user fairness. Each user will observe a different link rate and queuing delay. As a result, every user requires a separate target rate calculation at the \name router. 
The 3GPP cellular standard~\cite{3GPP} describes how scheduling information at the cellular base station can be used to calculate per-user link rates. This method is able to estimate capacity even if a given user is not backlogged at the base station, a key property for the target rate estimation in Equation~\eqref{eq:abctargetrule}.
\section{Discussion}
\label{s:discussion}

We discuss practical issues pertaining to ABC's deployment.

\smallskip
\noindent\textbf{Delayed Acks:} To support delayed ACKs, ABC uses byte counting at the sender; the sender increases/decreases its window by the new bytes ACKed. At the receiver, ABC uses the state machine from DCTCP~\cite{dctcp} for generating ACKs and echoing accel/brake marks. The receiver maintains the state of the last packet (accel or brake). Whenever the state changes, the receiver sends an ACK with the new state. If the receiver is in the same state after receiving $m$ packets (the number of ACKs to coalesce), then it sends a delayed ACK with the current state. Our TCP implementation and the experiments in \Sec{eval} use delayed ACKs with $m=2$.

\smallskip
\noindent\textbf{Lost ACKs:} ABC's window adjustment is robust to ACK losses. Consider a situation where the sender receives a fraction $p<1$ of the ACKs. If the accelerate fraction at the router is $f$, the current window of the sender is $w_{abc}$, then in the next RTT, the change in congestion window of the sender is $f p w_{abc} - (1 - f) p w_{abc} = (2f - 1) p w_{abc}$. As a result, lost ACKs only slow down the changes in the congestion window, but whether it increases or decreases doesn't depend on $p$.


\smallskip
\noindent\textbf{ABC routers don't change prior ECN marks:} ABC routers don't mark accel-brake on incoming packets that contain ECN marks set by an upstream non-ABC router. 
Since packets with ECN set can't convey accel-brake marks, they can slow down changes in $w_{abc}$ (similar to lost ACKs). In case the fraction of packets with ECN set is small, then, the slow down in changes to $w_{abc}$ will be small. If the fraction is large, then the non-ABC router is the likely bottleneck, and the sender will not use $w_{abc}$.

\smallskip
\noindent\textbf{ECN routers clobbering ABC marks:} An ECN router can overwrite accel-brake marks. 
The ABC sender will still track the non-ABC window, $w_{nonabc}$, but such marks can slow down adjustment to the ABC window, $w_{abc}$.

\smallskip
\noindent\textbf{ABC on fixed-rate links:}  ABC can also be deployed on fixed-rate links. On such links, its performance is similar to prior explicit schemes like XCP. 


\section{Evaluation}
\label{s:eval}

We evaluate \name by considering the following properties:
\begin{CompactEnumerate}
\item \textbf{Performance:} We measure \name's ability to achieve low delay and high throughput and compare \name to end-to-end schemes, AQM schemes, and explicit control schemes (\S\ref{ss:performance}). 
\item \textbf{Multiple Bottlenecks:} We test \name in scenarios with multiple \name bottlenecks and mixtures of \name and non-\name bottlenecks (\S\ref{ss:coexistence}).
\item \textbf{Fairness:} We evaluate \name's fairness while competing against other \name and non-\name flows (\S\ref{ss:flow_fairness}).
\item \textbf{Additional Considerations:} We evaluate how \name performs with application-limited flows and different network delays. We also demonstrate \name's impact on a real application's performance (\S\ref{ss:other_results}).
\end{CompactEnumerate}

\begin{figure*}[t]
    \centering
    \begin{subfigure}[tbh]{0.32\textwidth}
        \includegraphics[width=\textwidth]{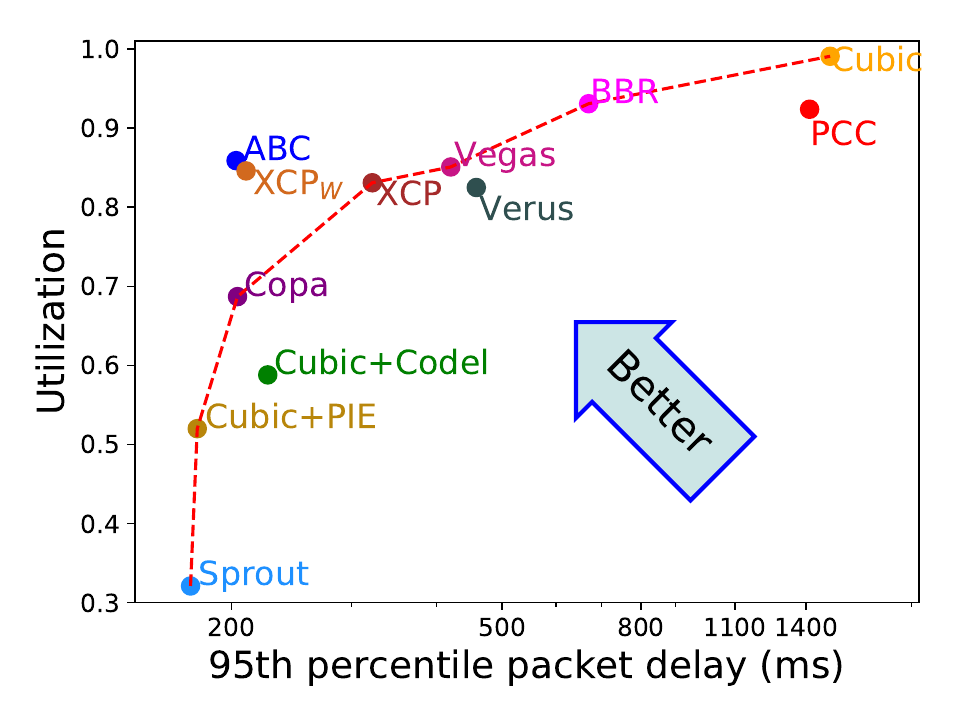}
        \vspace{-7mm}
        \caption{Downlink}
        \label{fig:througput-delay:down}
    \end{subfigure}
    \begin{subfigure}[tbh]{0.32\textwidth}
        \includegraphics[width=\textwidth]{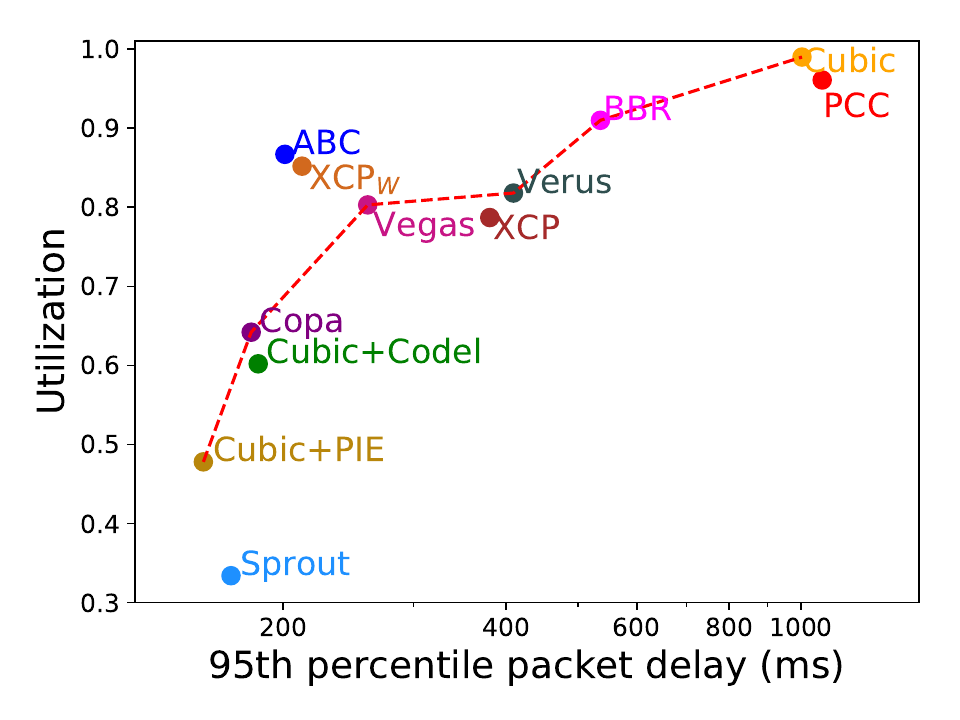}
        \vspace{-7mm}
        \caption{Uplink}
        \label{fig:througput-delay:up}
    \end{subfigure}
    \begin{subfigure}[tbh]{0.32\textwidth}
        \includegraphics[width=\textwidth]{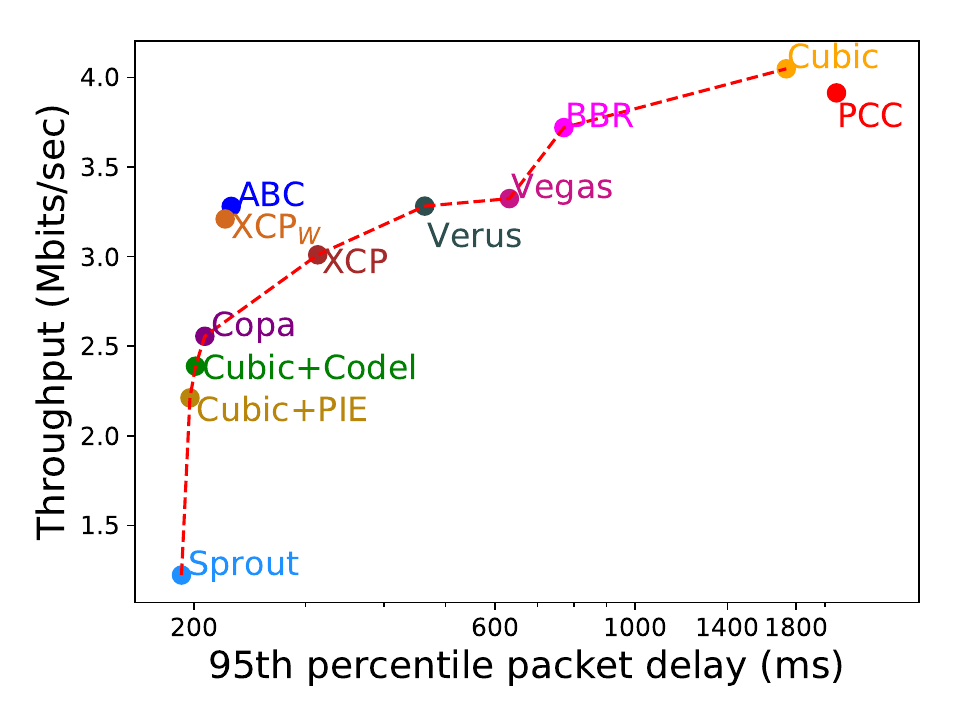}
        \vspace{-7mm}
        \caption{Uplink+Downlink}
        \label{fig:througput-delay:up+down}
    \end{subfigure}
    \vspace{-3mm}
    \caption{\small {\bf ABC vs. previous schemes on three Verizon cellular network traces ---} In each case, ABC outperforms all other schemes and sits well outside the Pareto frontier of previous schemes (denoted by the dashed lines). }
    \label{fig:througput-delay}
    \vspace{-6mm}
\end{figure*}

\subsection{Prototype \name Implementation}
\label{s:prototype implementation}
\smallskip
\noindent {\bf \name transport:} We implemented \name endpoints in Linux as kernel modules using the pluggable TCP API. 

\smallskip
\noindent{\bf \name router:} We implemented \name as a Linux queuing discipline (qdisc) kernel module using OpenWrt, an open source operating system for embedded networked devices~\cite{openwrt}. We used a NETGEAR WNDR 3800 router configured to 802.11n. We note that our implementation is portable as OpenWrt is supported on many other commodity Wi-Fi routers.

\name's WiFi link rate estimation exploits the inner workings of the MAC 802.11n protocol, and thus requires fine-grained values at this layer. In particular, the \name qdisc must know A-MPDU sizes, Block ACK receive times, and packet transmission bitrates. These values are not natively exposed to Linux router qdiscs, and instead are only available at the network driver. To bridge this gap, we modified the router to log the relevant MAC layer data in the cross-layer socket buffer data structure (skb) that it already maintains per packet.

\subsection{Experimental Setup}
\label{ss:setup}

We evaluated \name in both Wi-Fi and cellular network settings. For Wi-Fi, experiments we used a live Wi-Fi network and the \name router described in \Sec{prototype implementation}. For cellular settings, we use Mahimahi~\cite{mahimahi} to emulate multiple cellular networks (Verizon LTE, AT\&T, and TMobile). Mahimahi's emulation uses packet delivery traces (separate for uplink and downlink) that were captured directly on those networks, and thus include outages (highlighting \name's ability to handle ACK losses).



We compare \name to end-to-end protocols designed for cellular networks (Sprout~\cite{sprout} and Verus~\cite{verus}), loss-based end-to-end protocols both with and without AQM (Cubic~\cite{cubic}, Cubic+Codel~\cite{codel}, and Cubic+PIE~\cite{pie}), recently-proposed end-to-end protocols (BBR~\cite{bbr}, Copa~\cite{copa}, PCC Vivace-Latency (referred as PCC))~\cite{vivace}), and TCP Vegas~\cite{vegas}), and explicit control protocols (XCP~\cite{xcp}, RCP~\cite{rcp} and VCP~\cite{VCP}). We used TCP kernel modules for \name, BBR, Cubic, PCC, and Vegas; for these schemes, we generated traffic using iperf~\cite{iperf}. For the end-to-end schemes that are not implemented as TCP kernel modules (i.e., Copa, Sprout, Verus), we used the UDP implementations provided by the authors. Lastly, for the explicit control protocols (i.e., XCP, RCP, and VCP), we used our own implementations as qdiscs with Mahimahi to ensure compatibility with our emulation setup. We used Mahimahi's support of Codel and Pie to evaluate AQM.


Our emulated cellular network experiments used a minimum RTT of 100 ms and a buffer size of 250 MTU-sized packets. Additionally, \name's target rate calculation (Equation~\eqref{eq:abctargetrule}) used $\eta=0.98$ and $\delta=133$ ms. Our Wi-Fi implementation uses the link rate estimator from \S\ref{s:linkrate}, while our emulated cellular network setup assumes the realistic scenario that \name's router has knowledge of the underlying link capacity~\cite{3GPP}. 

\subsection{Performance}
\label{ss:performance}

\noindent {\bf Cellular:} \Fig{througput-delay:down} and~\ref{fig:througput-delay:up} show the utilization and 95$^{th}$ percentile per packet delay that a single backlogged flow achieves using each aforementioned scheme on two Verizon LTE cellular link traces. \name exhibits a better (i.e., higher) throughput/delay tradeoff than all prior schemes. In particular, \name sits well outside the Pareto frontier of the existing schemes, which represents the prior schemes that achieve higher throughput or lower delay than any other prior schemes.

Further analysis of \Fig{througput-delay:down}~and~\ref{fig:througput-delay:up} reveals that Cubic+Codel, Cubic+PIE, Copa, and Sprout are all able to achieve low delays that are comparable to \name. However, these schemes heavily underutilize the link. The reason is that, though these schemes are able to infer and react to queue buildups in a way that reduces delays, they lack a way of quickly inferring increases in link capacity (a common occurrence on time-varying wireless links), leading to underutilization. In contrast, schemes like BBR, Cubic, and PCC are able to rapidly saturate the network (achieving high utilization), but these schemes also quickly fill buffers and thus suffer from high queuing delays. Unlike these prior schemes, \name is able to quickly react to \textit{both} increases and decreases in available link capacity, enabling high throughput and low delays.

\begin{figure}[t]
     \centering
     \begin{subfigure}[tbh]{\columnwidth}
     \includegraphics[width=1.0\columnwidth]{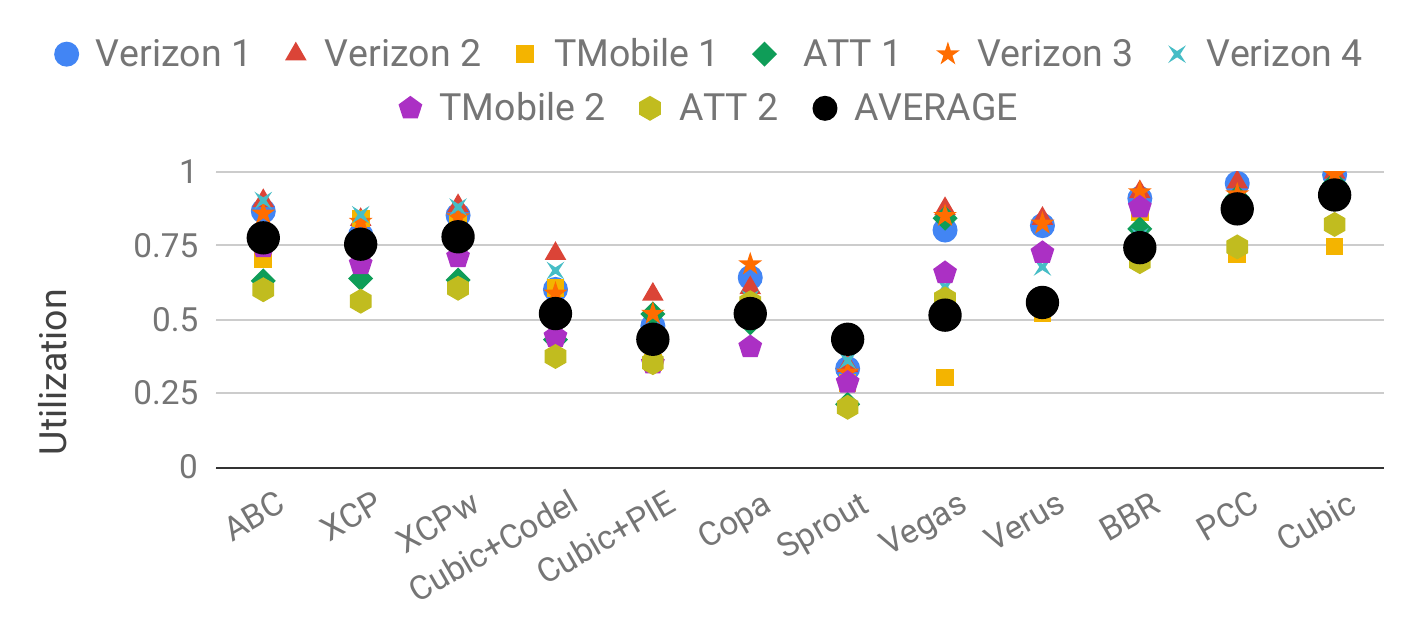}
    \vspace{-8mm}
    \subcaption{Utilization}
    \vspace{-1mm}
    \label{fig:aggregate-statistics:utilization}
    \end{subfigure}
    
    \begin{subfigure}[tbh]{\columnwidth}
    \includegraphics[width=1.0\columnwidth]{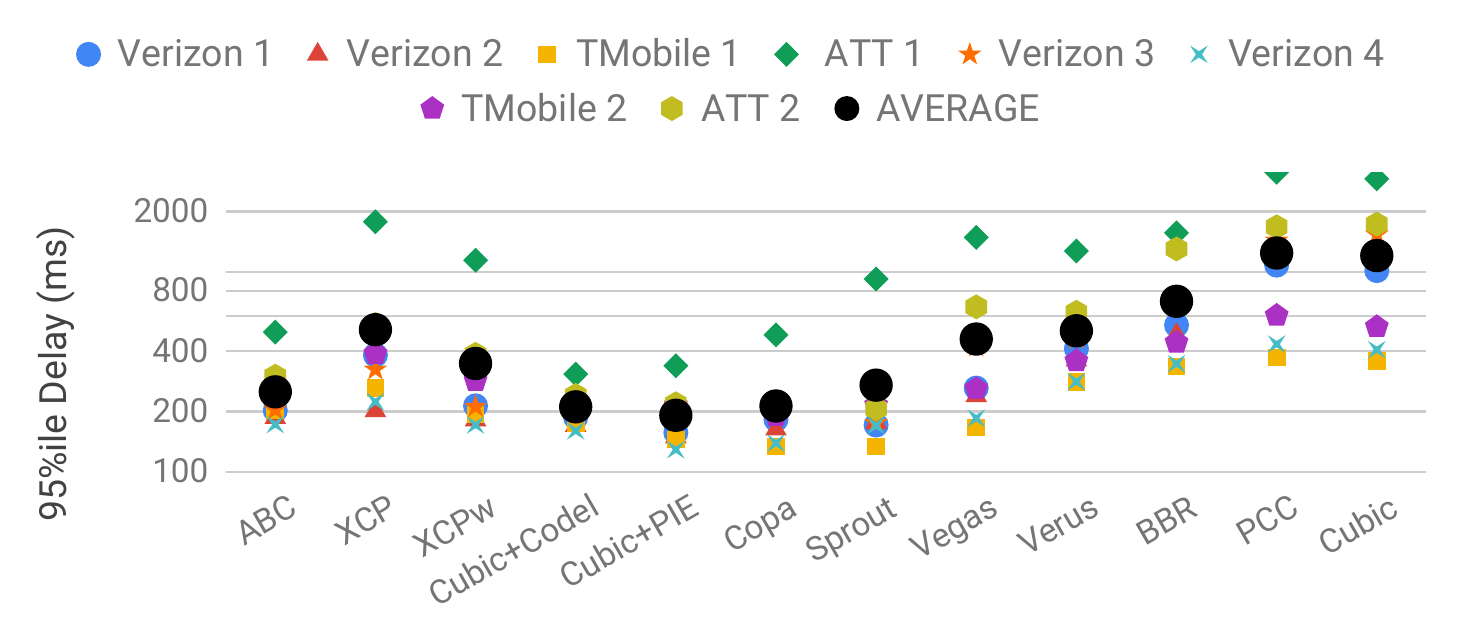}
    \vspace{-6mm}
    \subcaption{$95^{th}$ percentile per-packet delay}
    \label{fig:aggregate-statistics:delay}
    \end{subfigure}
    \vspace{-3mm}
    \caption{\small {\bf 95$^{th}$ percentile per-packet delay across 8 cellular link traces ---} On average, ABC achieves similar  delays  and  50\%  higher  utilization  than  Copa  and Cubic+Codel.  PCC  and  Cubic  achieve  slightly  higher throughput   than   ABC,   but   incur 380\% higher  95$^{th}$ percentile  delay than \name.}
    \label{fig:aggregate-statistics}
    \vspace{-6.5mm}
 \end{figure}

We observed similar trends across a larger set of 8 different cellular network traces (\Fig{aggregate-statistics}). \cut{\name is able to achieve higher throughput and lower delay than Sprout, Vegas, and Verus.} \name achieves 50\% higher throughput than Cubic+Codel and Copa, while only incurring 17\% higher $95^{th}$ percentile packet delays. PCC and Cubic achieve slightly higher link utilization values than \name (12\%, and 18\%, respectively), but incur significantly higher per-packet delays than \name (394\%, and 382\%, respectively). Finally, compared to BBR, Verus, and Sprout, \name achieves higher link utilization (4\%, 39\%, and 79\%, respectively). BBR and Verus incur higher delays (183\% and 100\%, respectively) than \name. Appendix~\ref{app:abc-throughput-delay} shows  mean packet delay over the same conditions, and shows the
same trends.

\smallskip
\noindent{\bf Comparison with Explicit Protocols:} \Fig{througput-delay} and~\ref{fig:aggregate-statistics} also show that \name outperforms the explicit control protocol, XCP, despite not using multi-bit per-packet feedback as XCP does. For XCP we used $\alpha=0.55$ and $\beta=0.4$, the highest permissible stable values that achieve the fastest possible link rate convergence. XCP achieves similar average throughput to \name, but with 105\% higher 95$^{th}$ percentile delays. This performance discrepancy can be attributed to the fact that \name's control rule is better suited for the link rate variations in wireless networks. In particular, unlike \name which updates its feedback on every packet, XCP computes aggregate feedback values ($\phi$) only once per RTT and may thus take an entire RTT to inform a sender to reduce its window. To overcome this, we also considered an improved version of XCP that recomputes aggregate feedback on each packet based on the rate and delay measurements from the past RTT; we refer to this version as XCP$_{w}$ (short for XCP wireless). As shown in \Fig{througput-delay} and \Fig{aggregate-statistics}, XCP$_{w}$ reduces delay compared to XCP, but still incurs 40\% higher 95$^{th}$ percentile delays (averaged across traces) than \name.
 We also compared with two other explicit schemes, RCP and VCP, and found that \name consistently outperformed both, achieving 20\% more utilization on average. (Appendix~\ref{app:explicit}).


 \begin{figure}[t]
     \centering
     \begin{subfigure}[t]{0.9\columnwidth}
    \includegraphics[width=1.0\columnwidth]{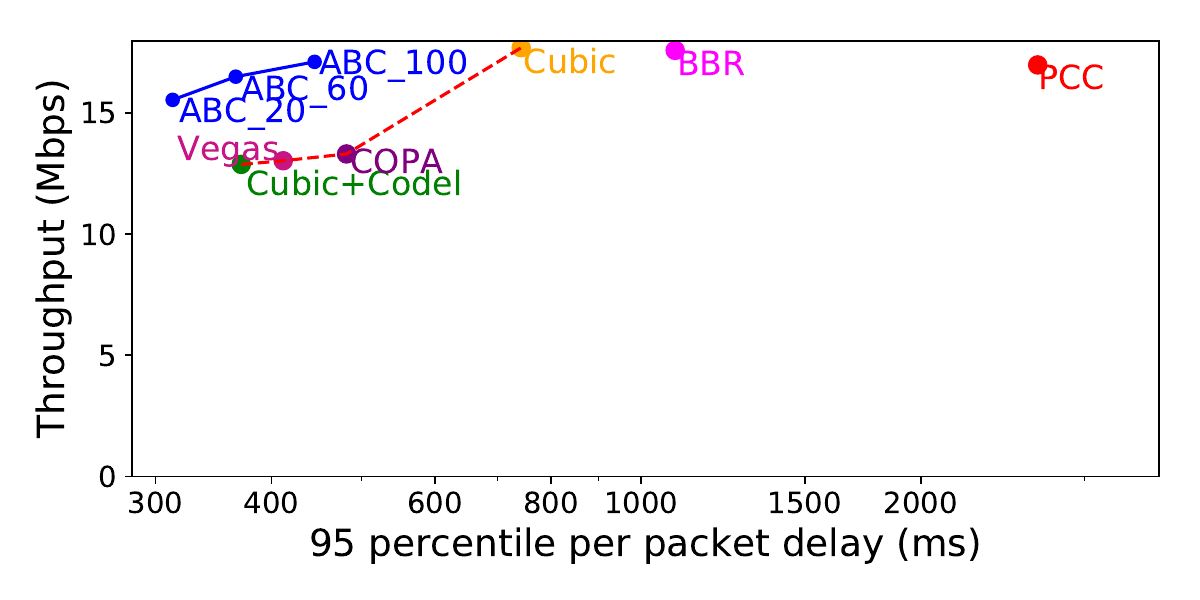}
    \vspace{-7.5mm}
    \subcaption{Single user}
    \label{fig:wifi:single}
    \vspace{-0.75mm}
    \end{subfigure}
    \begin{subfigure}[t]{0.9\columnwidth}
    \includegraphics[width=1.0\columnwidth]{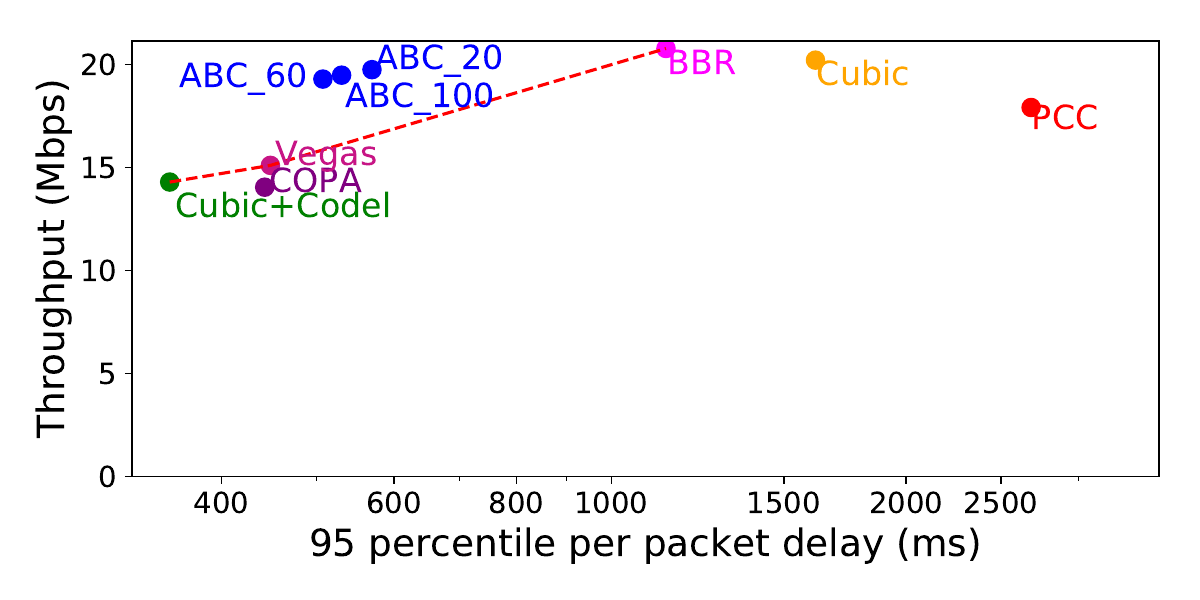}
    \vspace{-7mm}
    \subcaption{Two users, shared queue}
    \label{fig:wifi:multiple}
    \end{subfigure}
    \vspace{-3mm}
    \caption{\small {\bf Throughout and mean delay on Wi-Fi ---} For the multi-user scenario, we report the sum of achieved throughputs and the average of observed $95^{th}$ percentile delay across both users. We consider three versions of \name (denoted \name\_*) for different delay thresholds. All versions of \name outperform all prior schemes and sit outside the pareto frontier.}
    \label{fig:wifi}
    \vspace{-6mm}
 \end{figure}

\noindent {\bf Wi-Fi:} We performed similar evaluations on a live Wi-Fi link, considering both single and multi-user scenarios. {\color{black} We connect senders to a WiFi router via Ethernet. Each sender transmits data through the WiFi router to one receiver. All receivers' packets share the same FIFO queue at the router.} In this experiment, we excluded Verus and Sprout as they are designed specifically for cellular networks. To mimic common Wi-Fi usage scenarios where endpoints can move and create variations in signal-to-noise ratios (and thus bitrates), we varied the Wi-Fi router's bitrate selections by varying the MCS index using the Linux \texttt{iw} utility; we alternated the MCS index between values of 1 and 7 every 2 seconds. In Appendix~\ref{fig:app:wifi_brownian}, we also list results for an experiment where we model MCS index variations as Brownian motion---results show the same trends as described below. This experiment was performed in a crowded computer lab with contention from other Wi-Fi networks. We report average performance values across three, 45 second runs. We considered three different \name delay threshold ($d_{t}$) values of 20 ms, 60 ms, and 100 ms; note that increasing \name's delay threshold will increase both observed throughput and RTT values.


\Fig{wifi} shows the throughput and $95^{th}$ percentile per-packet delay for each protocol\cut{, as measured using the \texttt{tcptrace} Linux utility}. For the multi-user scenario, we report the sum of achieved throughputs and the average $95^{th}$ percentile delay across all users. In both the single and multi-user scenarios, \name achieves a better throughput/delay tradeoff than all prior schemes, and falls well outside the Pareto frontier for those schemes. In the single user scenario, the \name configuration with $d_{t}=100$ ms achieves up to 29\% higher throughput than Cubic+Codel, Copa and Vegas. Though PCC-Vivace, Cubic and BBR achieve slightly higher throughput (4\%) than this \name configuration, their delay values are considerably higher (67\%-6$\times$). The multi-user scenario showed similar results. For instance, \name achieves 38\%, 41\% and 31\% higher average throughput than Cubic+Codel, Copa and Vegas, respectively.

\subsection{Coexistence with Various Bottlenecks}
\label{ss:coexistence}

\textbf{Coexistence with \name bottlenecks:} 
\Fig{througput-delay:up+down} compares \name and prior protocols on a network path with two cellular links. In this scenario, \name{} tracks the bottleneck link rate and achieves a better throughput/delay tradeoff than prior schemes, and again sits well outside the Pareto frontier.\cut{ for those schemes.}

 \begin{figure}[t]
     \centering
     \includegraphics[width=0.9\columnwidth]{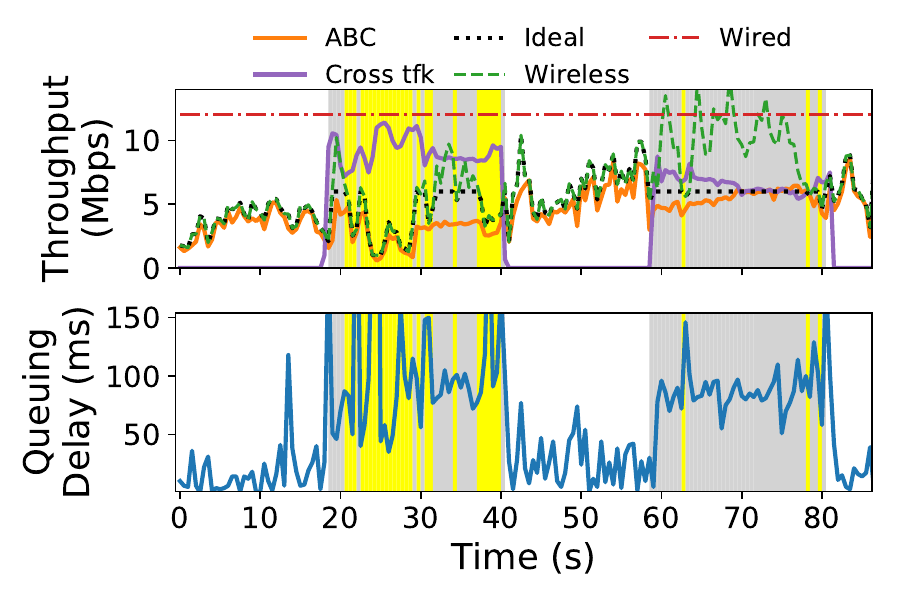}
     \vspace{-4.5mm}
     \caption{\small {\bf Coexistence with non-\name bottlenecks ---} \name tracks the ideal rate closely (fair share) and reduces queuing delays in the absence of cross traffic (white region). }
     \label{fig:ts_droptail_cell_visualization}
     \vspace{-6.5mm}
\end{figure}

\smallskip
\noindent{\bf Coexistence with non-\name bottlenecks:} 
\Fig{ts_droptail_cell_visualization} illustrates throughput and queuing delay values for an \name flow traversing a network path with both an emulated wireless link and an emulated 12 Mbits/s fixed rate (wired) link. The wireless link runs \name, while the wired link operates a droptail buffer. \name shares the wired link with on-off cubic cross traffic. In the absence of cross traffic (white region), the wireless link is always the bottleneck. However, with cross traffic (yellow and grey regions), due to contention, the wired link can become the bottleneck. In this case, \name's fair share on the wired link is half of the link's capacity (i.e., 6 Mbit/s).
If the wireless link rate is lower than the fair share on the wired link (yellow region), the wireless link remains the bottleneck; otherwise, the wired link becomes the bottleneck (grey region).

The black dashed line in the top graph represents the ideal fair throughput \cut{(i.e., that of the bottleneck link)} for the \name flow throughout the experiment. As shown, in all regions, \name is able to track the ideal rate closely, even as the bottleneck shifts. In the absence of cross traffic, \name achieves low delays while maintaining high link utilization. With cross traffic, \name appropriately tracks the wireless link rate (yellow region) or achieves its fair share of the wired link (grey region) like Cubic. In the former cross traffic scenario, increased queuing delays are due to congestion caused by the Cubic flow on the wired link. Further, deviations from the ideal rate in the latter cross traffic scenario can be attributed to the fact that the \name flow is running as Cubic, which in itself takes time to converge to the fair share ~\cite{cubic}.



\subsection{Fairness among \name and non-\name flows}
\label{ss:flow_fairness}
\label{s:flow_fairness}

\textbf{Coexistence among \name flows:} We simultaneously run multiple \name flows on a fixed 24 Mbits/s link. We varied the number of competing flows from 2 to 32 (each run was 60 s). In each case, the Jain Fairness Index~\cite{jain1999throughput} was within 5\% from the ideal fairness value of 1, highlighting \name's ability to ensure fairness.

\begin{figure}[t]
    \centering
    \includegraphics[width=0.9\columnwidth]{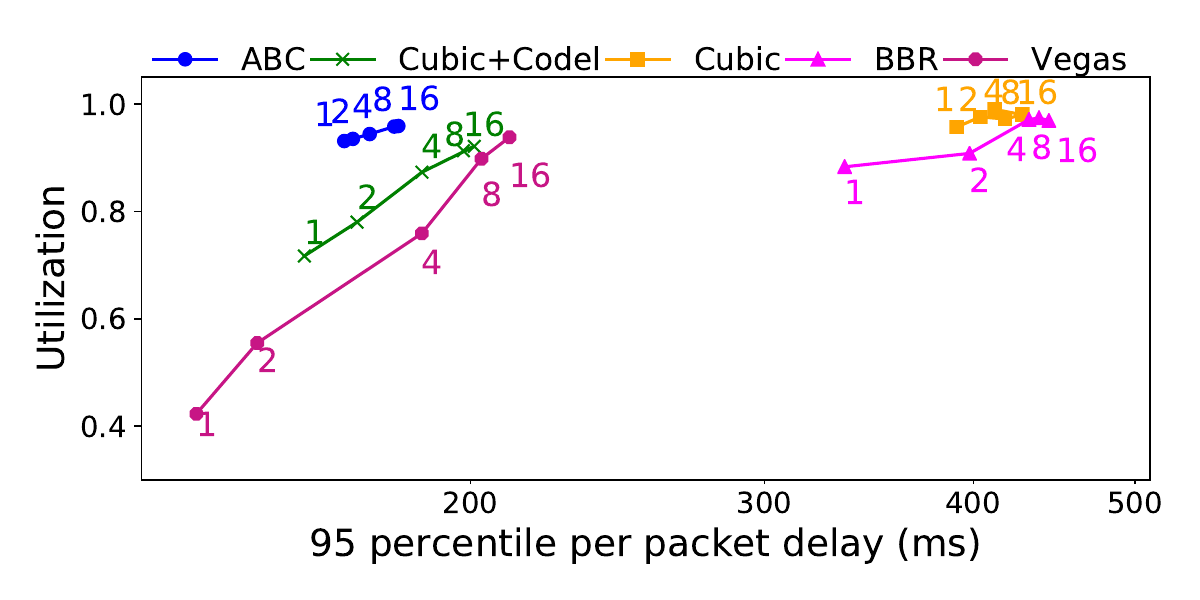}
    \vspace{-4.5mm}
    \caption{\small {\bf Coexistence among ABC flows ---} ABC achieves similar aggregate utilization and delay irrespective of the number of connections. \name outperforms all previous schemes.}
    \label{fig:multiple_flow}
    \vspace{-5mm}
 \end{figure}

\Fig{multiple_flow} shows the aggregate utilization and delay for concurrent flows (all flows running the same scheme) competing on a Verizon cellular link. We varied the number of competing flows from 1 to 16. ABC achieves similar aggregate utilization and delay across all scenarios, and, outperforms all other schemes. For all the schemes, the utilization and delay increase when the number of flows increases. For ABC, this increase can be attributed to the additional packets that result from additive increase (1 packet per RTT per flow). For other schemes, this increase is because multiple flows in aggregate ramp-up their rates faster than a single flow. 

\smallskip
\noindent\textbf{RTT Unfairness:}  We simultaneously ran 2 ABC flows on a 24 Mbits wired bottleneck. We varied the RTT of flow 1 from 20ms to 120ms. RTT of flow 2 was fixed to 20ms. \Fig{rtt_unfairness} shows the ratio of the average throughput of these 2 flows (average throughput of flow 2 / flow 1, across 5 runs) against the ratio of their RTTs (RTT of flow 1 / flow 2). Increasing the RTT ratio increases the throughput ratio almost linearly and the throughput is inversely proportional to the RTT. Thus, the unfairness is similar to  existing protocols like Cubic.

Next, we simultaneously ran 6 ABC flows. The RTT of the flows vary from 20ms to 120ms. Table \ref{table:rtt_unfairness} shows the RTT and the average throughput across 5 runs. Flows with higher RTTs have lower throughput. However, note that the flow with the highest RTT (120ms) still achieves $\sim$35 \% of the throughput as flow with the lowest RTT (20ms). 


\begin{table}[]
\small
\begin{minipage}[b]{0.45\linewidth}
\centering
\includegraphics[width=33mm,height=33mm]{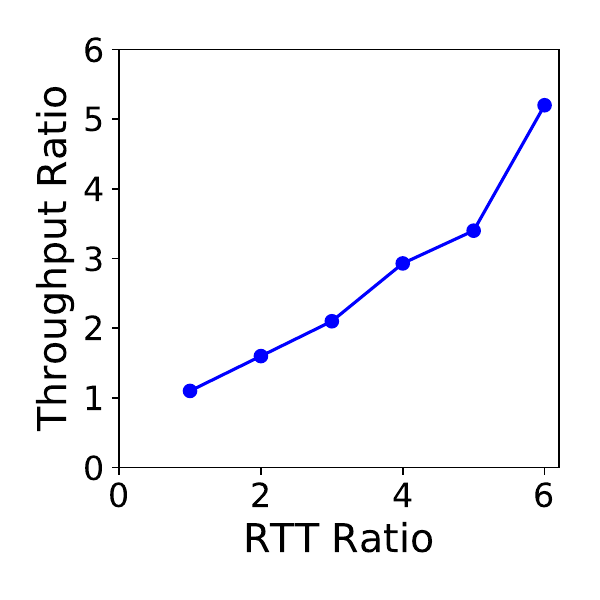}
\vspace{-4mm}
\captionof{figure}{\textbf{\small {RTT unfairness}}}
\label{fig:rtt_unfairness}
\end{minipage}
\begin{minipage}[b]{0.45\linewidth}
\centering
\begin{tabular}{ c | c }

    RTT (ms) & Tput (Mbps)\\ \hline \hline
    20 & 6.62 \\ \hline
    40 & 4.94 \\ \hline
    60 & 4.27 \\ \hline
    80 & 3.0 \\ \hline
    100 & 2.75 \\ \hline
    120 & 2.40 \\ 

   \end{tabular}
   \vspace{-1mm}
    \caption{\textbf{\small RTT unfairness}}
    \label{table:rtt_unfairness}
\end{minipage}\hfill

\vspace{-4.5mm}
\end{table}



\begin{figure}
    \centering
    \begin{subfigure}[t]{0.22\textwidth}
        \includegraphics[width=\textwidth]{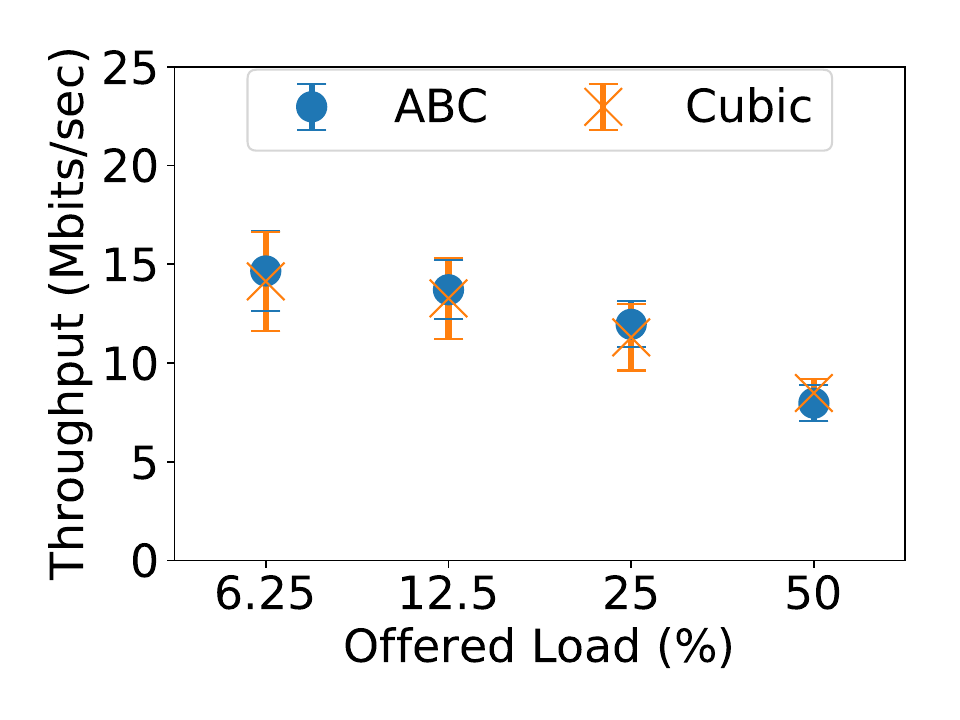}
        \vspace{-7mm}
        \caption{ABC}
        \label{fig:non-abc-flow:abc}
    \end{subfigure}
    \begin{subfigure}[t]{0.22\textwidth}
        \includegraphics[width=\textwidth]{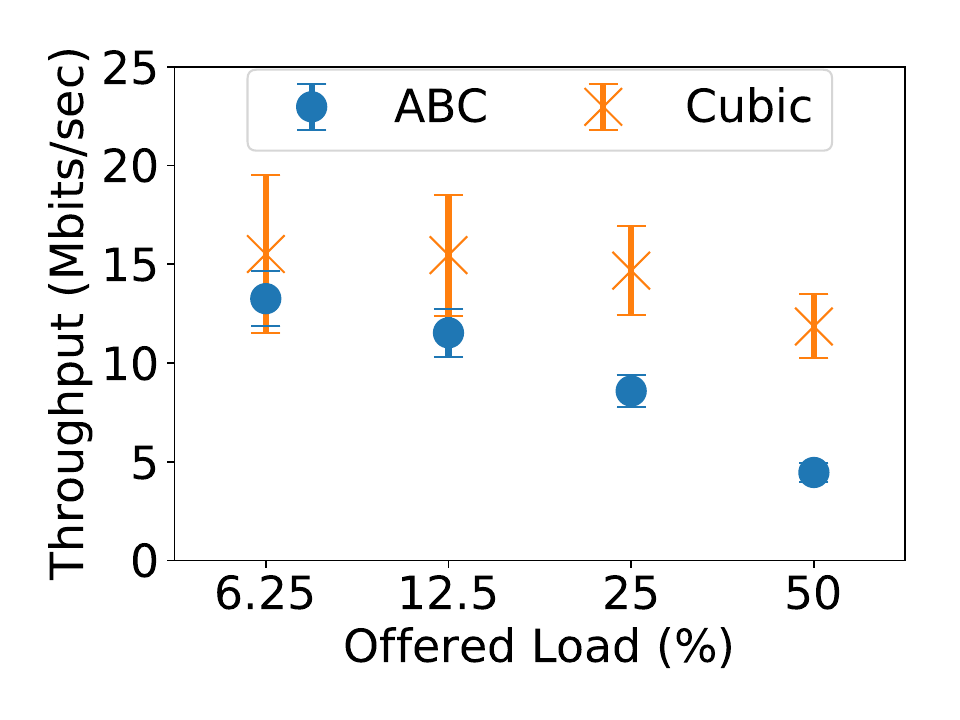}
        \vspace{-7mm}
        \caption{RCP's Zombie List}
        \label{fig:non-abc-flow:rcp}
    \end{subfigure}
    \vspace{-3mm}
    \caption{\small{\bf Coexistence with non-\name flows ---} Across all scenarios, the standard deviation for \name flows is small and the flows are fair to each other. Compared to RCP's Zombie List strategy, \name's max-min allocation provides better fairness between \name and non-\name flows. With \name's strategy, the difference in average throughput of \name and Cubic flows is under 5\%.}
    \label{fig:non-abc-flow}
    \vspace{-6mm}
\end{figure}

\smallskip
\noindent{\bf Coexistence with non-\name flows:} We consider a scenario where 3 \name and 3 non-\name (in this case, Cubic) long-lived flows share the same 96 Mbits/s bottleneck link. In addition, we create varying numbers of non-\name short flows (each of size 10 KB) with Poisson flow arrival times to offer a fixed average load. We vary the offered load values, and report results across 10 runs (40 seconds each). We compare \name's strategy to coexist with non-\name flows to RPC's Zombie list approach (\S\ref{s:fairness}).

\Fig{non-abc-flow} shows the mean and standard deviation of throughput for long-lived \name and Cubic flows. As shown in \Fig{non-abc-flow:abc}, \name's coexistence strategy allows \name and Cubic flows to fairly share the bottleneck link across all offered load values. Specifically, the difference in average throughput between the \name and Cubic flows is under 5\%. In contrast, \Fig{non-abc-flow:rcp} shows that RCP's coexistence strategy gives higher priority to Cubic flows. This discrepancy increases as the offered load increases, with Cubic flows achieving 17-165\% higher throughput than \name flows. The reason, as discussed in ~\Sec{fairness}, is that long-lived Cubic flows receive higher throughput than the average throughput that RCP estimates for Cubic flows. This leads to unfairness because RCP attempts to match average throughput for each scheme.\Fig{non-abc-flow} also shows that the standard deviation of \name flows is small (under 10\%) across all scenarios. This implies that in each run of the experiment, the throughput for each of the three concurrent \name flows is close to each other, implying fairness across \name flows. Importantly, the standard deviation values for \name are smaller than those for Cubic. Thus, \name flows converge to fairness faster than Cubic flows do.

\subsection{Additional Results}
\label{ss:other_results}

\cut{\begin{figure}[t]
     \centering
     \begin{subfigure}[t]{\columnwidth}
    \includegraphics[width=1.0\columnwidth]{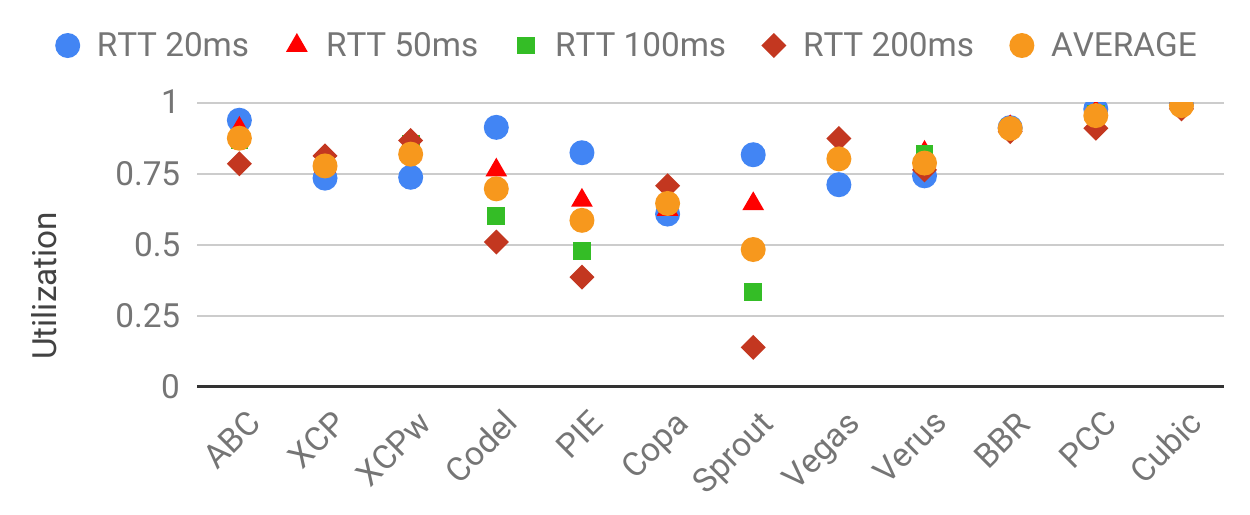}
    \vspace{-6mm}
    \subcaption{Utilization}
    \end{subfigure}
    \begin{subfigure}[t]{\columnwidth}
    \includegraphics[width=1.0\columnwidth]{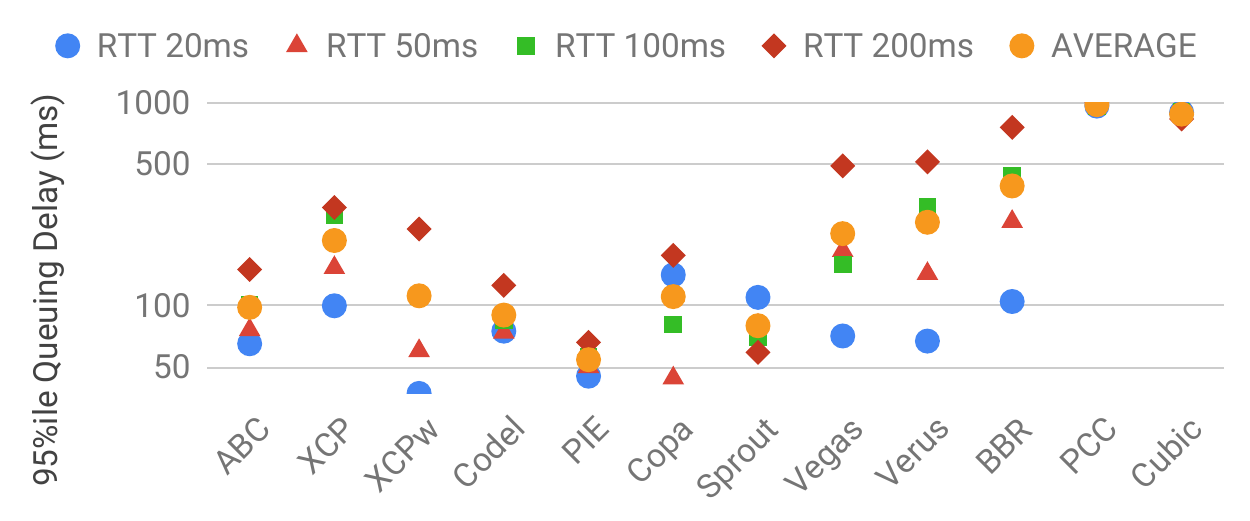} \vspace{-6mm}
    \subcaption{ per-packet queuing delay}
    \end{subfigure}
    \vspace{-3mm}
    \caption{\small {\bf Impact of propagation delay on performance ---} On a Verizon cellular network trace with different propagation delays, \name achieves a better throughput/delay tradeoff than all other schemes.}
    \label{fig:rtt}
    \vspace{-4mm}
 \end{figure}}


\noindent{\bf \name's sensitivity to network latency:} Thus far, our emulation experiments have considered fixed minimum RTT values of 100 ms. To evaluate the impact that propagation delay on \name's performance, we repeated the experiment from \Fig{aggregate-statistics} on the RTT values of 20 ms, 50 ms, 100 ms, and 200 ms. Across all RTTs, \name outperforms all prior schemes, again achieving a more desirable throughput/latency trade off (see Appendix ~\ref{app:other}). \cut{\name's benefits persist even though schemes like Cubic+Codel and Cubic+PIE actually improve with increasing propagation delays. Performance with these schemes improves because bandwidth delay products increase, making Cubic's additive increase more aggressive (improving link utilization).} 

\smallskip
\noindent{\bf Application-limited flows:} We created a single long-lived \name flow that shared a cellular link with 200 application-limited \name flows that send traffic at an aggregate of 1 Mbit/s. Despite the fact that the application-limited flows do not have traffic to properly respond to \name's feedback, the \name flows (in aggregate) still achieve low queuing delays and high link utilization. See Appendix ~\ref{app:other} for details.

\smallskip
\noindent{\bf Perfect future capacity knowledge:} We considered a variant of \name, PK-\name, which knows an entire emulated link trace in advance. This experiment reflects the possibility of resource allocation predictions at cellular base stations. Rather than using an estimate of the current link rate to compute a target rate (as \name does), PK-\name uses the expected link rate 1 RTT in the future. 
On the same setup as \Fig{througput-delay:up}, PK-\name reduces 95$^{th}$ percentile per-packet-delays from 97 ms to 28 ms, compared to \name, while achieving similar utilization ($\sim$90\%). 

\smallskip
\noindent{\bf \name's improvement on real applications:} We evaluated \name's improvement for real user-facing applications on a multiplayer interactive game, Slither.io~\cite{slither}. We loaded Slither.io using a Google Chrome browser which ran inside an emulated cellular link with a background backlogged flow. We considered three schemes for the backlogged flow: Cubic, Cubic+Codel, and \name. Cubic fully utilizes the link, but adds excessive queuing delays hindering gameplay. Cubic+Codel reduces queuing delays (improving user experience in the game), but underutilizes the link. Only \name is able to achieve both high link utilization for the backlogged flow and low queuing delays for the game. A video demo of this experiment can be viewed at \url{https://youtu.be/Dauq-tfJmyU}.

\noindent{\bf Impact of $\eta$:} This parameter presents a trade-off between throughput and delay. Increasing $\eta$ increases the throughput but at the cost of additional delay (see \App{eta}).

\cut{\smallskip
\noindent\textbf{Other experiments:} We also evaluated 1) The impact on \name's performance when sharing the link with application-limited flows. 2) \name's sensitivity to network latency (or base RTT). Please see Appendix ~\ref{app:other} for details. In particular, \name outperforms all previous schemes across all RTT values.}

\section{Related Work}
\label{s:related}

Several prior works have proposed using LTE infrastructure to infer the underlying link capacity~\cite{xie2015pistream, lu2015cqic, jain2015mobile}. CQIC~\cite{lu2015cqic} and piStream~\cite{xie2015pistream} use physical layer information at the receiver \cut{(e.g., allocated physical resource blocks)} to estimate link capacity. However, these approaches have several limitations that lead to inaccurate estimates. CQIC's estimation approach considers historical resource usage (not the available physical resources)~\cite{xie2015pistream}, while piStream's technique relies on second-level video segment downloads and thus does not account for the short timescale variations in link rate required for per-packet congestion control. These inaccuracies stem from the opacity of the base station's resource allocation process at the receiver. \name circumvents these issues by accurately estimating link capacity directly at the base station.

In VCP~\cite{VCP}, router classifies congestion as low, medium, or high, and signals the sender to either perform a multiplicative increase, additive increase, or multiplicative decrease in response. Unlike an \name sender, which reacts to ACKs individually, VCP senders act once per RTT. This coarse-grained update  limits VCP's effectiveness on time-varying wireless paths. For instance, it can take 12 RTTs to double the window. VCP is also incompatible with ECN, making it difficult to deploy. 

In BMCC~\cite{bmcc, bmcc_new}, a router uses ADPM~\cite{adpm} to send link load information to the receiver on ECN bits, relying on TCP options to relay the feedback from the receiver to the sender. 
MTG proposed modifying cellular base stations to communicate the link rate explicitly using a new TCP option~\cite{jain2015mobile}. Both approaches do not work with IPSec encryption~\cite{seo2005security}, and such packet modifications trigger the risk of packets being dropped silently by middleboxes~\cite{honda2011still}. Moreover, unlike \name, MTG does not ensure fairness among multiple flows for a user, while BMCC has the same problem with non-BMCC flows~\cite{bmcc_new, bmcc++}. 

XCP-b~\cite{xcpb} is a variant of XCP designed for wireless links with unknown capacity. XCP-b routers use the queue size to determine the feedback. When the queue is backlogged, the XCP-b router calculates spare capacity using the change in queue size and uses the same control rule as XCP. When the queue goes to zero, XCP-b cannot estimate spare capacity, and resorts to a blind fixed additive increase. Such blind increase can cause both under-utilization and increased delays (\Sec{case_explicit}.)

Although several prior schemes (XCP, RCP, VCP, BMCC, XCP-b) attempt to match the current enqueue rate to the capacity, none match the future enqueue rate to the capacity, and so do not perform as well as ABC on time-varying links.



\section{Conclusion}
\label{s:conclusion}

This paper presented a simple new explicit congestion control protocol for time-varying wireless links called ABC. ABC routers use a single bit to mark each packet with ``accelerate'' or ``brake'', which causes senders to slightly increase or decrease their congestion windows. Routers use this succinct feedback to quickly guide senders towards a desired target rate. ABC outperforms the best existing explicit flow control scheme, XCP, but unlike XCP, \name does not require modifications to packet formats or user devices, making it simpler to deploy. ABC is also incrementally deployable: \name can operate correctly with multiple \name and non-\name bottlenecks, and can fairly coexist with \name and non-\name traffic sharing the same bottleneck link. We evaluated \name using a WiFi router implementation and trace-driven emulation of cellular links. ABC achieves 30-40\% higher throughput than Cubic+Codel for similar delays, and 2.2$\times$ lower delays than BBR on a Wi-Fi path. On cellular network paths, ABC achieves 50\% higher throughput than Cubic+Codel.

\noindent\textbf{Acknowledgments.}
We thank Anirudh Sivaraman, Peter Iannucci, and Srinivas Narayana for useful discussions. We are grateful to the anonymous reviewers and our shepherd Jitendra Padhye for their feedback and useful comments.  This work was supported in part by DARPA under Contract No. HR001117C0048, and NSF grants CNS-1407470, CNS-1751009, and CNS-1617702.

\bibliographystyle{abbrv}
\bibliography{abc}



\clearpage
\appendix
\cut{\section{Comparison of \name{} feedback based on enqueue vs. dequeue rate}
\label{app:deque-vs-enque}

The \name router calculates accelerate fraction $f(t)$ based on dequeue rate (Equation ~\ref{eq:f}). We also evaluate the performance when the feedback is calculated using enqueue rate ($eq(t)$) instead. In this case, $f(t) = \min\Big\{\frac{1}{2} \cdot \frac{tr(t)}{eq(t)}, 1 \Big\}$. \Fig{illustration} shows that using enqueue rate for feedback increases the 95$^{th}$ percentile queuing delay by 2 $\times$. 

\begin{figure}[t]
    \centering
    \begin{subfigure}[t]{0.23\textwidth}
        \includegraphics[width=\textwidth]{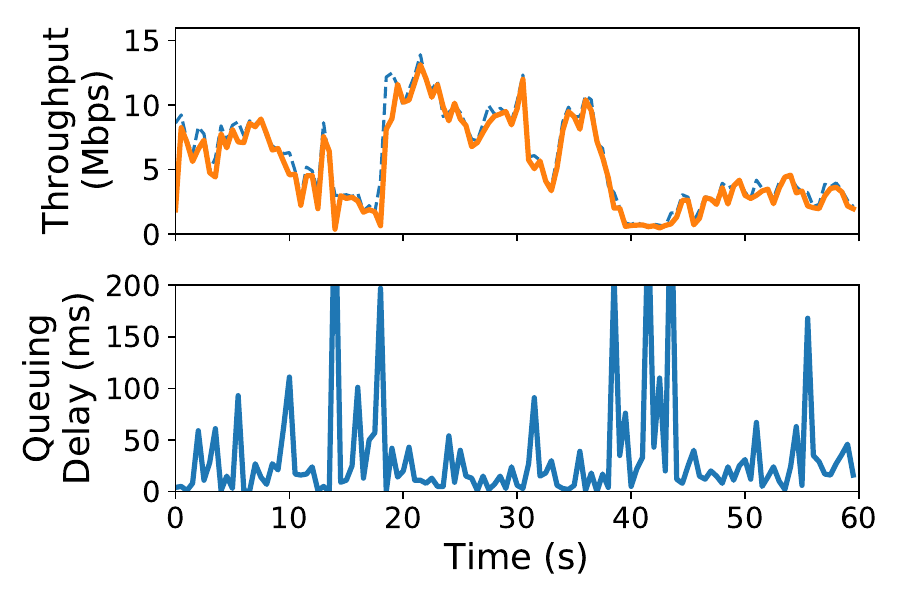}
        \vspace{-6mm}
        \caption{Dequeue}
        \label{fig:illustration:dq}
    \end{subfigure}
    \begin{subfigure}[t]{0.23\textwidth}
        \includegraphics[width=\textwidth]{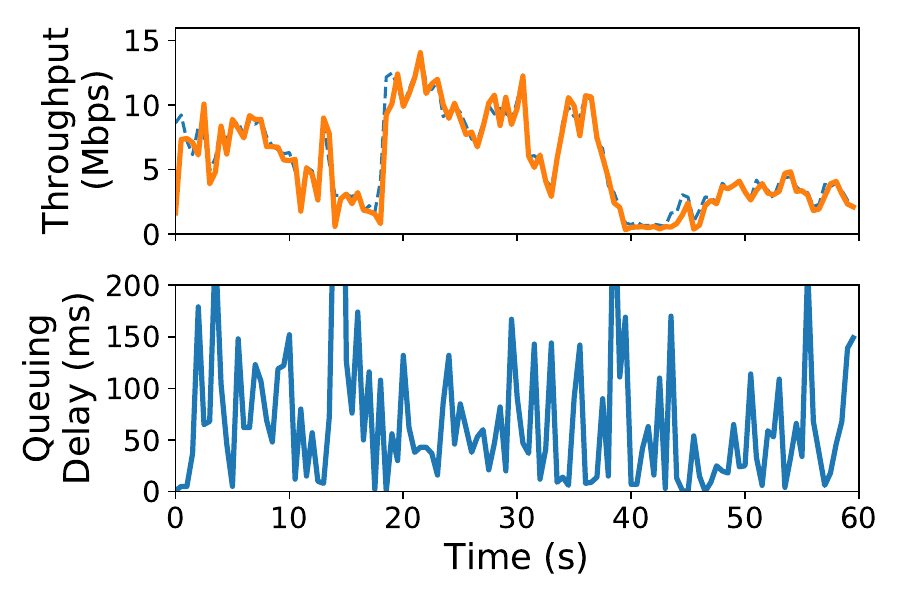}
        \vspace{-6mm}
        \caption{Enqueue}
        \label{fig:illustration:eq}
    \end{subfigure}
    \vspace{-3mm}
    \caption{\small{\bf Feedback ---} Calculating $f(t)$ based on enqueue rate increases delay.}
    \label{fig:illustration}
    \vspace{-5mm}
\end{figure}}

\section{BBR overestimates the sending rate}
\label{app:comp_bbr}

\begin{figure}[t]
    \centering
    \begin{subfigure}[t]{0.23\textwidth}
        \includegraphics[width=\textwidth]{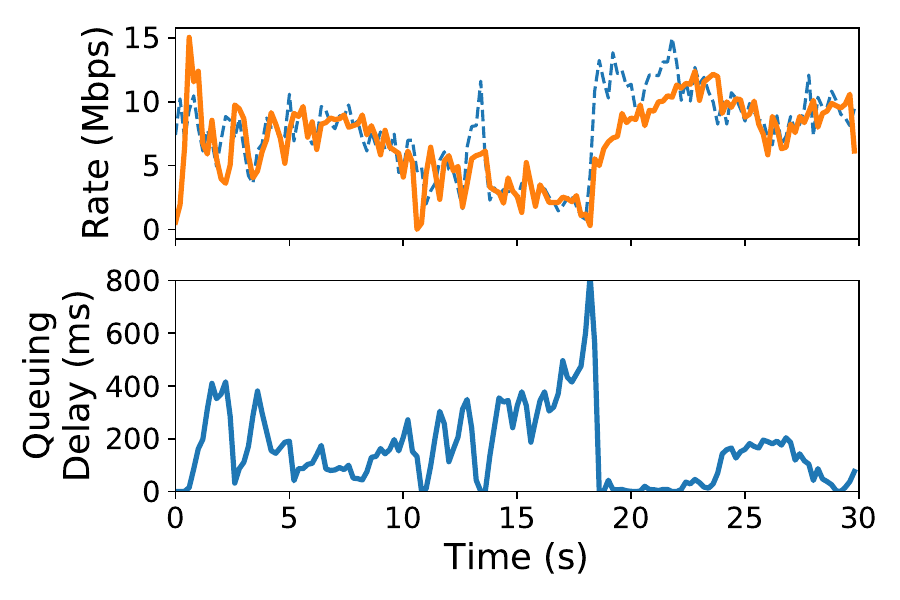}
        \vspace{-6mm}
        \caption{BBR}
        \label{fig:comp_bbr:bbr}
    \end{subfigure}
    \begin{subfigure}[t]{0.23\textwidth}
        \includegraphics[width=\textwidth]{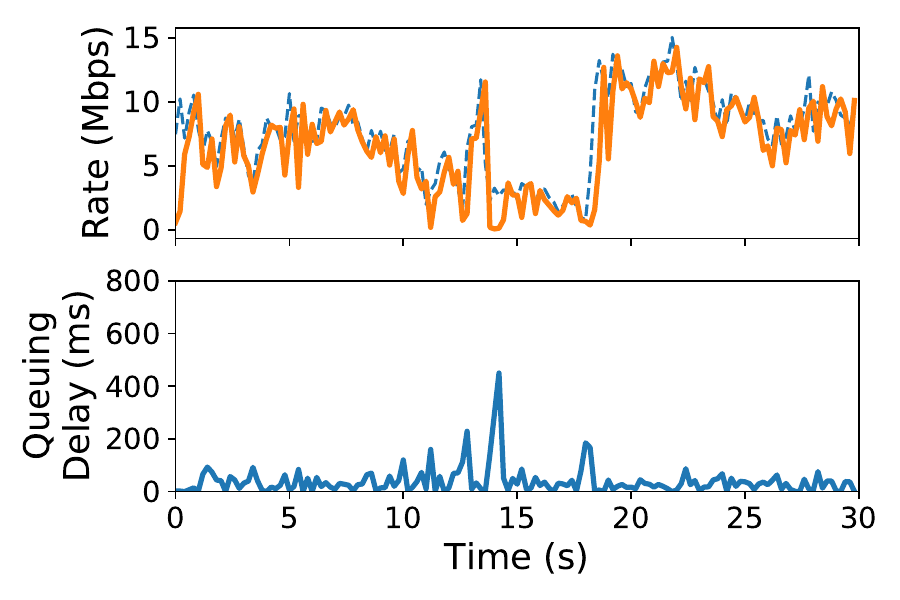}
        \vspace{-6mm}
        \caption{ABC}
        \label{fig:comp_bbr:abc}
    \end{subfigure}
    \vspace{-3mm}
    \caption{\small{\bf Comparison with BBR ---} BBR overshoots the link capacity, causing excessive queuing. Same setup as \Fig{motivation}.}
    \label{fig:comp_bbr}
    \vspace{-4mm}
\end{figure}

{\color{black} \Fig{comp_bbr} shows the throughput and queuing delay of BBR on a Verizon cellular trace. BBR periodically increases its rate in short pulses, and frequently overshoots the link capacity with variable-bandwidth links, causing excessive queuing.}

 \begin{figure}[t]
     \centering
     \includegraphics[width=0.78\columnwidth]{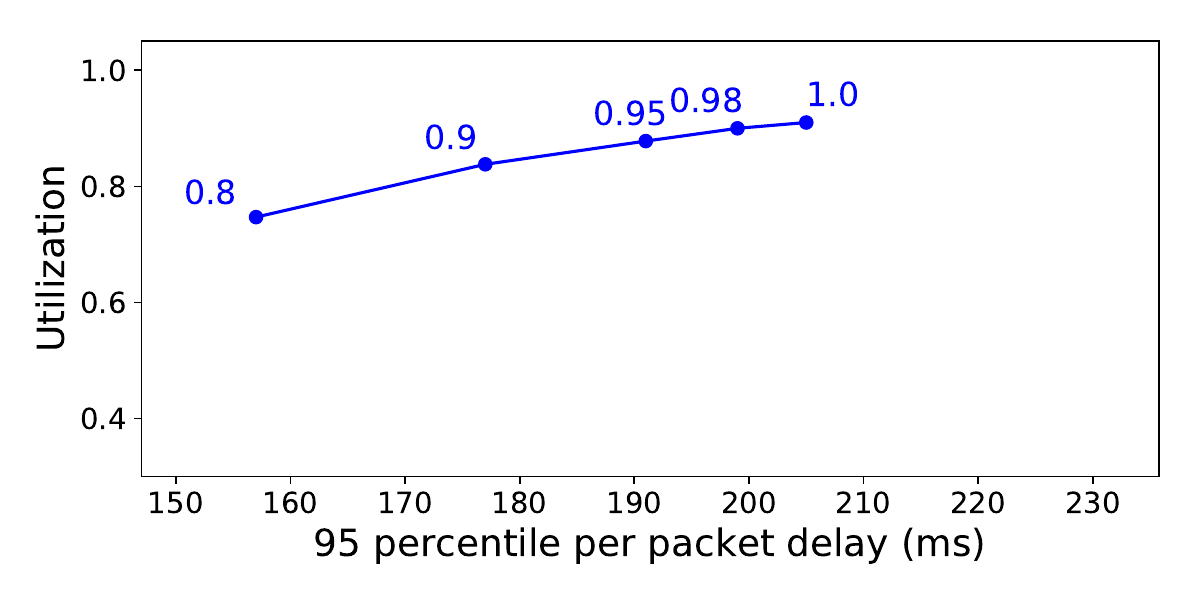}
     \vspace{-4.5mm}
     \caption{\small {\bf Impact of $\eta$ ---} Performance of ABC with various values of $\eta$ (target utilization). $\eta$ presents a trade-off between throughput and delay. Same setup as \Fig{motivation}.}
     \label{fig:eta}
     \vspace{-6.5mm}
\end{figure}

\section{Impact of $\eta$}
\label{app:eta}
{\color{black} \Fig{eta} shows the performance of ABC with various values of $\eta$ on a Verizon cellular trace. Increasing $\eta$ increases the link utilization, but, also increases the delay. Thus, $\eta$ presents a trade-off between throughput and delay.}

\section{Stability Analysis}
\label{app:stability}
This section establishes the stability bounds for \name's control algorithm (Theorem~\ref{thm:stablity}). 

\noindent \textbf{Model:} Consider a single \name link, traversed by $N$ \name flows. Let $\mu(t)$ be the link capacity at time $t$. As $\mu(t)$ can be time-varying, we define stability as follows. Suppose that at some time $t_0$, $\mu(t)$ stops changing, i.e., for $t>t_0$ $\mu(t) = \mu$ for some constant $\mu$. We aim to derive conditions on \name's parameters which guarantee that the aggregate rate of the senders and the queue size at the routers will converge to certain fixed-point values (to be determined) as $t \to \infty$.   

Let $\tau$ be the common round-trip propagation delay on the path for all users. For additive increase (\S\ref{s:fairness}), assume that each sender increases its congestion window by 1 every $l$ seconds. Let $f(t)$ be the fraction of packets marked accelerate, and, $cr(t)$ be the dequeue rate at the \name router at time $t$. Let $\tau_r$ be time it takes accel-brake marks leaving the \name router to reach the sender. Assuming that there are no queues other than at the \name router, $\tau_r$ will be the sum of the propagation delay between  \name router and the receiver and the propagation delay between receiver and the senders. The aggregate incoming rate of ACKs across all the senders at time $t$, $R(t)$, will be equal to the dequeue rate at the router at time $t-\tau_r$:
 \begin{align}
    R(t) \;&=\;  cr(t-\tau_r).
\end{align}
In response to an accelerate, a sender will send 2 packets, and, for a brake, a sender won't send anything. In addition to responding to accel-brakes, each sender will also send an additional packet every $l$ seconds (because of AI). Therefore, the aggregate sending rate for all senders at time $t$, $S(t)$, will be 
 \begin{align}
    S(t) \;&=\;  R(t) \cdot 2 \cdot f (t-\tau_r)+ \frac{N}{l}\nonumber\\
    \;&=\; 2 cr(t-\tau_r) f(t-\tau_r) + \frac{N}{l}.
\end{align}
Substituting $f(t-\tau_r)$ from Equation~\eqref{eq:f}, we get 
\begin{align}
    S(t) \;&=\; tr(t-\tau_r) + \frac{N}{l}.
\end{align}
Let $\tau_f$ be the propagation delay between a sender and the \name router, and $eq(t)$ be the enqueue rate at the router at time $t$. Then $eq(t)$ is given by
 \begin{align}
    eq(t) \;&=\;  S(t-\tau_f)\nonumber\\
    \;&=\; tr(t-(\tau_r+\tau_f)) + \frac{N}{l}\nonumber\\
     \;&=\; tr(t-\tau) + \frac{N}{l}.
    \label{eq:stability:eq}
\end{align}
Here, $\tau = \tau_r + \tau_f$ is the round-trip propagation delay. 

Let $q(t)$ be the queue size, and, $x(t)$ be the queuing delay at time $t$: 
\begin{align}
    x(t) \;&=\;  \frac{q(t)}{\mu}.\nonumber
\end{align}   
Ignoring the boundary conditions for simplicity ($q(t)$ must be $\geq$ 0), the queue length has the following dynamics: 
\begin{align}
 \dot{q}(t) \;&=\; eq(t) - \mu\nonumber\\
  \;&=\; tr(t-\tau) + \frac{N}{l} - \mu \nonumber\\
  \;&=\; \left((\eta-1) \cdot \mu + \frac{N}{l}\right) - \frac{\mu}{\delta}(x(t-\tau)-d_t)^{+}, \nonumber
\end{align}
where in the last step we have used Equation~\eqref{eq:abctargetrule}. Therefore the dynamics of $x(t)$ can be described by:
 \begin{align}
 \dot{x}(t) \;&=\; \left((\eta-1) + \frac{N}{\mu \cdot l}\right) - \frac{1}{\delta}(x(t-\tau)-d_t)^{+}  \nonumber\\
 \;&=\; A - \frac{1}{\delta}(x(t-\tau)-d_t)^{+},
 \label{eq:stability:system}
\end{align}
where $A = \left((\eta-1) + \frac{N}{\mu \cdot l}\right)$, and, A is a constant given a fixed number of flows $N$. The delay-differential equation in Equation~\eqref{eq:stability:system} captures the behavior of the entire system. We use it to analyze the behavior of the queuing delay, $x(t)$, which in turn informs the dynamics of the target rate, $tr(t)$, and enqueue rate, $eq(t)$, using Equations~\eqref{eq:abctargetrule} and~\eqref{eq:stability:eq} respectively. 

\noindent \textbf{Stability:} For stability, we consider two possible scenarios 1) $A < 0$, and 2) $A \geq 0$. We argue the stability in each case.

\noindent \textbf{Case 1: $A < 0 $}. In this case, the stability analysis is straightforward. The fixed point for queuing delay, $x^{*}$, is 0. From Equation~\eqref{eq:stability:system}, we get
\begin{align}
    \dot{x}(t) \;=\; A - \frac{1}{\delta}(x(t-\tau)-d_t)^{+}\;\leq\; A \;< 0.
    \label{eq:stability:case1}
\end{align}
The above equation implies that the queue delay will decrease at least as fast as $A$. Thus, the queue will go empty in a bounded amount of time. Once the queue is empty, it will remain empty forever, and the enqueue rate will converge to a fixed value. Using Equation~\eqref{eq:stability:eq}, the enqueue rate can will converge to
\begin{align}
    eq(t) \;&=\; tr(t-\tau) + \frac{N}{l}\nonumber\\
     \;&=\; \eta  \mu +\frac{N}{l} - \frac{\mu}{\delta}(x(t-\tau)-d_t)^{+} \nonumber\\
     \;&=\; \eta \mu + \frac{N}{l}\nonumber\\
     \;&=\; (1 + A) \mu.
\end{align}
Note that $\eta \mu < (1 + A) \mu < \mu$. Since both the enqueue rate and the queuing delay converge to fixed values, the system is stable for any value of $\delta$.

\smallskip
\noindent \textbf{Case 2: $A > 0 $}: The fixed point for the queuing delay in this case is $x^{*}= A \cdot \delta + d_t$. Let $\overset{\sim}{x}(t) = x(t) - x^*$ be the deviation of the queuing delay from its fixed point. Substituting in Equation~\eqref{eq:stability:system}, we get
\begin{align}
    \dot{\overset{\sim}{x}}(t) \;&=\; A - \frac{1}{\delta}(\overset{\sim}{x}(t-\tau) + A \cdot \delta)^{+}\nonumber\\
     \;&=\; - max(- A, \frac{1}{\delta}\overset{\sim}{x}(t-\tau)) \nonumber\\
     \;&=\; -g(\overset{\sim}{x}(t-\tau)),
    \label{eq:stability:approx}
\end{align}
where $g(u) = \max(-A, \frac{1}{\delta}u)$ and $A > 0$. 

In ~\cite{yorke1970asymptotic} (Corollary 3.1), Yorke established that delay-differential equations of this type are globally asymptotically stable (i.e., $\overset{\sim}{x}(t) \to 0$ as $t \to \infty$ irrespective of the initial condition), if the following conditions are met:
\begin{enumerate}
    \item \textbf{H$_{1}$:} g is continuous.
    \item \textbf{H$_{2}$:}  There exists some $\alpha$, s.t. $\alpha \cdot u^2 > ug(u) > 0$ for all $u \neq 0$.
    \item \textbf {H$_{3}$:} $\alpha \cdot \tau < \frac{3}{2}$.
\end{enumerate}

The function $g(\cdot)$ trivially satisfies \textbf{H$_{1}$}. \textbf{H$_2$} holds for any $\alpha \in (\frac{1}{\delta}, \infty)$. Therefore, there exists an $\alpha \in (\frac{1}{\delta}, \infty)$ that satisfies both \textbf{H$_2$} and \textbf{H$_3$} if  
\begin{align}
    \frac{1}{\delta} \cdot \tau < \frac{3}{2} \implies \delta > \frac{2}{3} \cdot \tau.
    \label{eq:stability:criterion}
\end{align}
This proves that \name's control rule is asymptotically stable if Equation~\eqref{eq:stability:criterion} holds. Having established that $x(t)$ converges to $x^* = A\cdot\delta + d_t$, we can again use Equation~\eqref{eq:stability:eq} to derive the fixed point for the enqueue rate:
\begin{align}
    eq(t) = \eta  \mu +\frac{N}{l} - \frac{\mu}{\delta}(x(t-\tau)-d_t)^{+} 
    \to \mu, 
\end{align}
as $t \to \infty$.

Note while, we proved stability assuming that the feedback delay $\tau$ is a constant and the same value for all the senders, the proof works even if the senders have different time-varying feedback delays (see Corollary 3.2 in \cite{yorke1970asymptotic}). The modified stability criterion in this case is $\delta > \frac{2}{3} \cdot \tau^*$, where $\tau^*$ is the maximum feedback delay across all senders.

\section{Wi-Fi Evaluation}
\begin{figure}[t]
    \centering
    \includegraphics[width=\columnwidth]{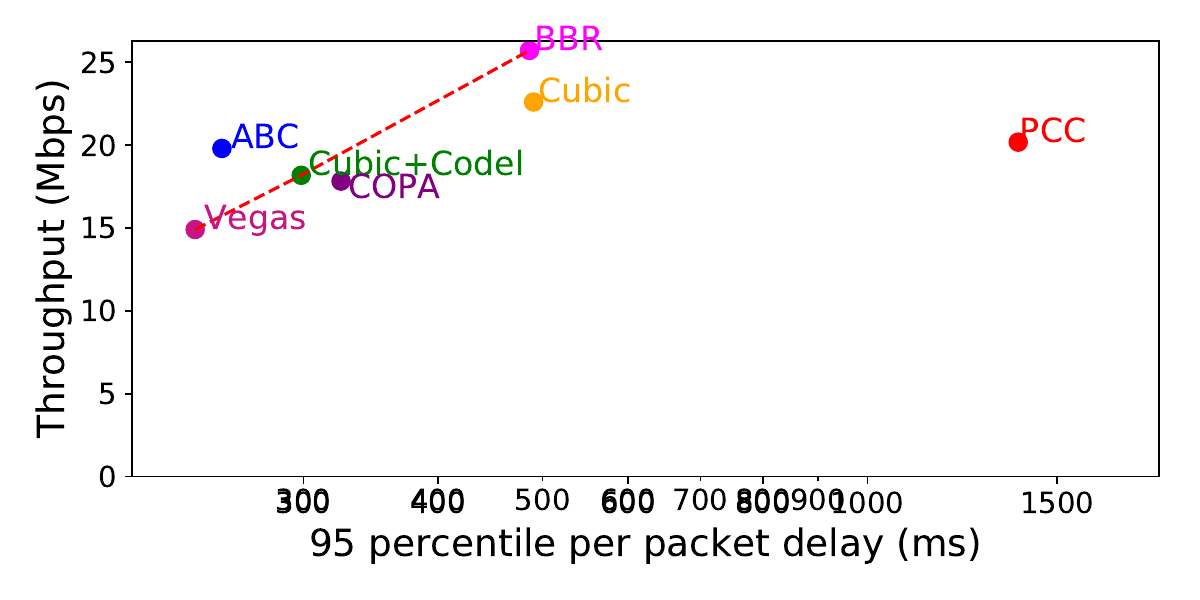}
    \caption{\small {\bf Throughput and $95^{th}$ percentile delay for a single user in WiFi ---} We model changes in MCS index as bownian motion, with values changing every 2 seconds. We limit the MCS index values to be between 3 and 7. \name outperforms all other schemes.}
    \label{fig:app:wifi_brownian}
\end{figure}

In this experiment we use the setup from \Fig{wifi:single}. To emulate movement of the receiver, we model changes in MCS index as brownian motion, with values changing every 2 seconds. \Fig{app:wifi_brownian} shows throughput and $95^{th}$ percentile per packet delay for a number of schemes. Again, \name outperforms all other schemes achieving better throughput and latency trade off.

\section{Low Delays and High Throughput}
\label{app:abc-throughput-delay}

\begin{figure}[t]
    \centering
    \includegraphics[width=\columnwidth]{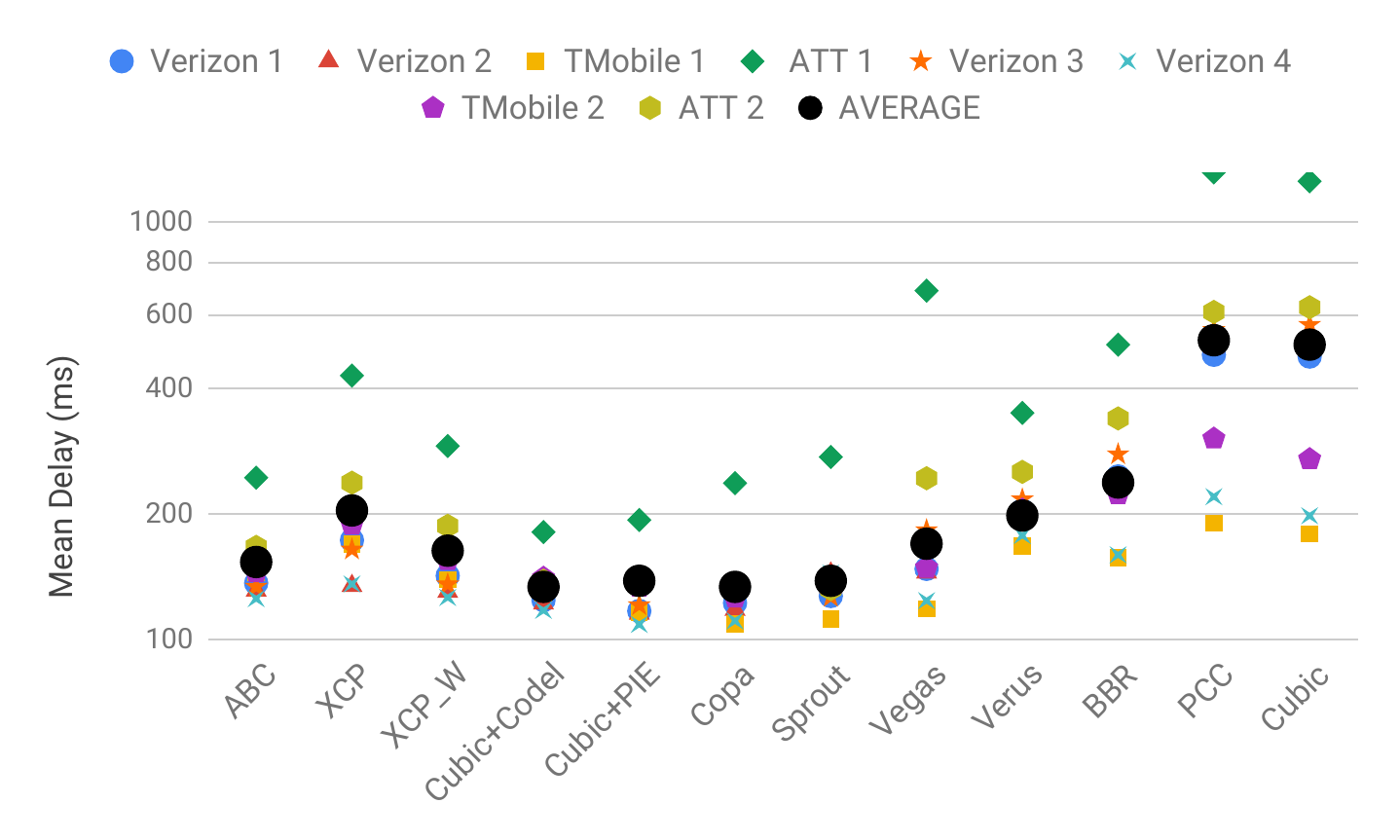}
    \caption{\small {\bf Utilization and mean per-packet delay across 8 different cellular network traces ---} On average, \name achieves similar delays and 50\% higher utilization than Copa and Cubic+Codel. BBR, PCC, and Cubic achieve slightly higher throughput than \name, but incur 70-240\% higher mean per-packet delays.}
    \label{fig:app:agg_95delay}
\end{figure}

\Fig{app:agg_95delay} shows the mean per packet delay achieved by various schemes in the experiment from \Fig{aggregate-statistics}. We observe the trend in mean delay is similar to that of 95$^{th}$ percentile delay ( \Fig{aggregate-statistics:delay}). \name{} achieves delays comparable to Cubic+Codel, Cubic+PIE and Copa. BBR, PCC Vivace-latency and Cubic incur 70-240\% higher mean delay than \name.

\section{\name vs Explicit Control Schemes}
\label{app:explicit}
\begin{figure}[t]
    \centering
    \begin{subfigure}[t]{0.23\textwidth}
        \includegraphics[width=\textwidth]{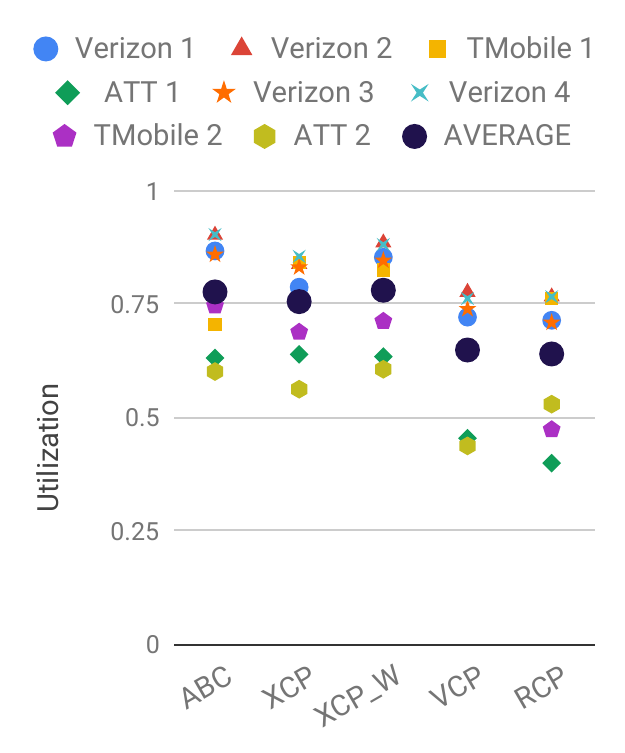}
        \caption{Utilization}
        \label{fig:explicit:utilization}
    \end{subfigure}
    \begin{subfigure}[t]{0.23\textwidth}
        \includegraphics[width=\textwidth]{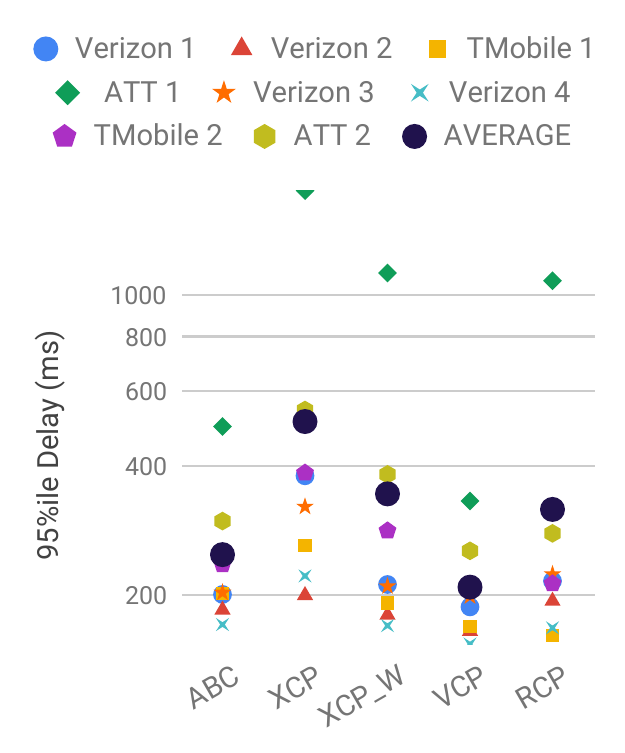}
        \caption{Delay}
        \label{fig:explicit:delay}
    \end{subfigure}
    \caption{\small{\bf \name vs explicit flow control ---} \name achieves similar utilization and $95^{th}$ percentile per-packet delay as XCP and XCP$_{w}$ across all traces. Compared to RCP and VCP, \name achieves 20\% more utilization.}
    \label{fig:explicit}
\end{figure}

\begin{figure*}
    \centering
    \begin{subfigure}[t]{0.3\textwidth}
        \includegraphics[width=\textwidth]{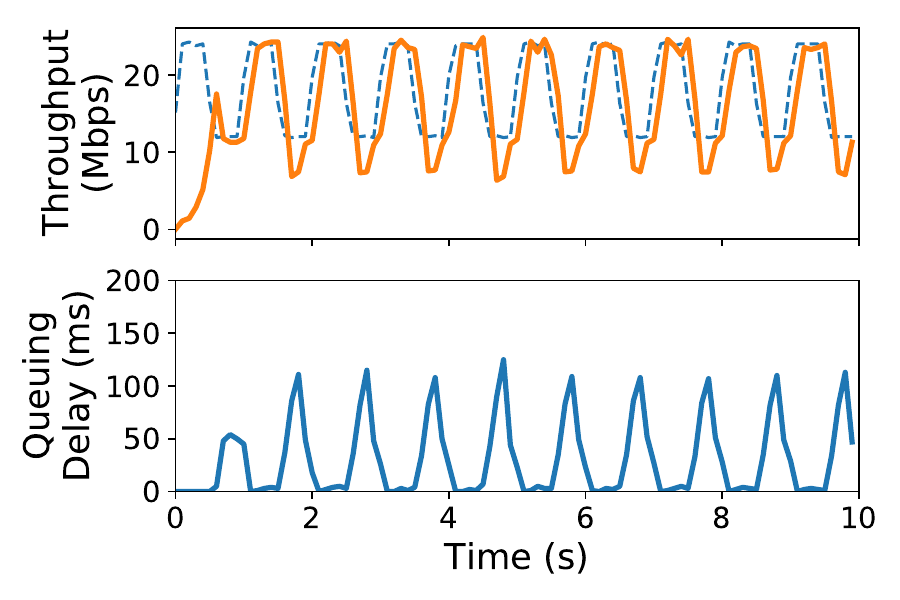}
        \caption{ABC}
        \label{fig:explicit:abc}
    \end{subfigure}
    \begin{subfigure}[t]{0.3\textwidth}
        \includegraphics[width=\textwidth]{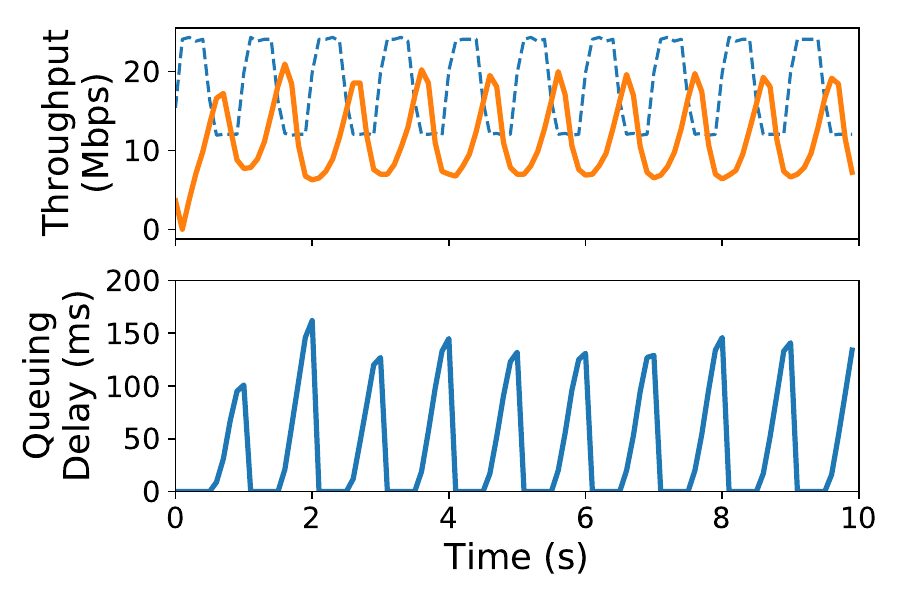}
        \caption{RCP}
        \label{fig:explicit:rcp}
    \end{subfigure}
    \begin{subfigure}[t]{0.3\textwidth}
        \includegraphics[width=\textwidth]{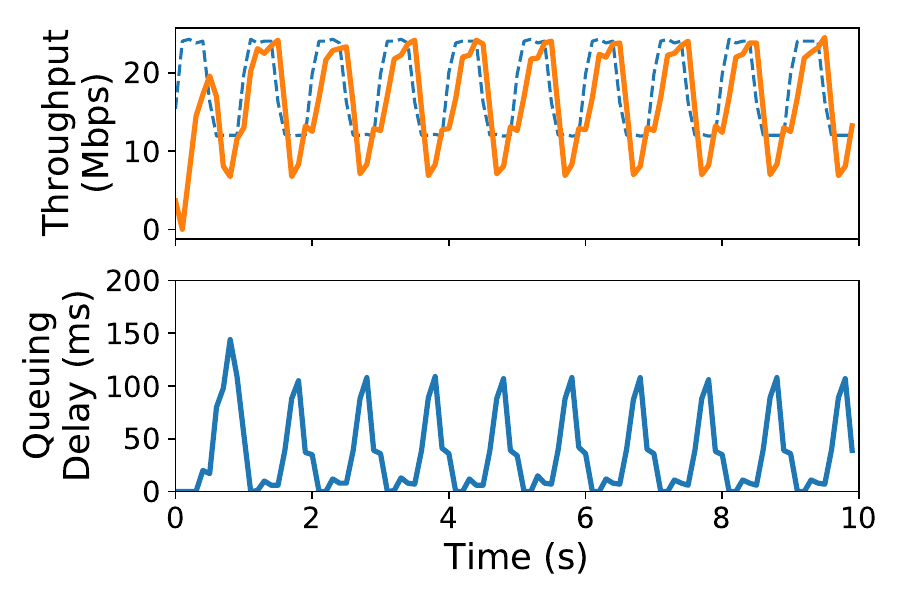}
        \caption{XCP$_{w}$}
        \label{fig:explicit:xcp}
    \end{subfigure}
    \caption{\small {\bf Time series for explicit schemes ---} We vary the link capacity every 500ms between two rates 12 Mbit/sec and 24 Mbit/sec.The dashed blue in the top graph represents bottleneck link capacity. \name and XCP$_{w}$ adapt quickly and accurately to the variations in bottleneck rate, achieving close to 100\% utilization. RCP is a rate base protocol and is inherently slower in reacting to congestion. When the link capacity drops, RCP takes time to drain queues and over reduces its rates, leading to under-utilization.}
    \label{fig:app:explicit}
\end{figure*}

In this section we compare \name's performance with explicit congestion control schemes. We consider XCP, VCP, RCP and our modified implementation of XCP (XCP$_{w}$). For XCP and XCP$_{w}$, we used constant values of $\alpha=0.55$ and $\beta=0.4$, which the authors note are the highest permissible stable values that achieve the fastest possible link rate convergence. For RCP and VCP, we used the author-specified parameter values of $\alpha=0.5$ and $\beta=0.25$, and $\alpha=1$, $\beta=0.875$ and $\kappa=0.25$, respectively. \Fig{explicit} shows utilizations and mean per packet delays achieved by each of these schemes over eight different cellular link traces. As shown, \name is able to achieve similar throughput as the best performing explicit flow control scheme, XCP$_{w}$, without using multibit per-packet feedback. We note that XCP$_{w}$'s 95$^{th}$ percentile per-packet delays are 40\% higher than \name's. \name is also able to outperform RCP and VCP. Specifically, \name achieves 20\% higher utilization than RCP. This improvement stems from the fact that RCP is a rate based protocol (not a window based protocol)---by signaling rates, RCP is slower to react to link rate fluctuations (Figure~\ref{fig:app:explicit} illustrates this behavior). \name also achieves 20\% higher throughput than VCP, while incurring slightly higher delays. VCP also signals multiplicative-increase/multiplicative-decrease to the sender. But unlike \name, the multiplicative increase/decrease constants are fixed. This coarse grained feedback limits VCP's performance on time varying links.

\Fig{app:explicit} shows performance of \name, RCP and XCP$_{w}$ on a simple time varying link. The capacity alternated between 12 Mbit/sec and 24 Mbit/sec every 500 milliseconds. \name and XCP$_{w}$ adapt quickly and accurately to the variations in bottleneck rate, achieving close to 100\% utilization. RCP is a rate base protocol and is inherently slower in reacting to congestion. When the link capacity drops, RCP takes time to drain queues and over reduces its rates, leading to under-utilization.

\section{Other experiments}
\label{app:other}
\noindent\textbf{Application limited flows} 

\begin{figure}[t]
    \centering
        \includegraphics[width=0.8\columnwidth,height=1.5in]{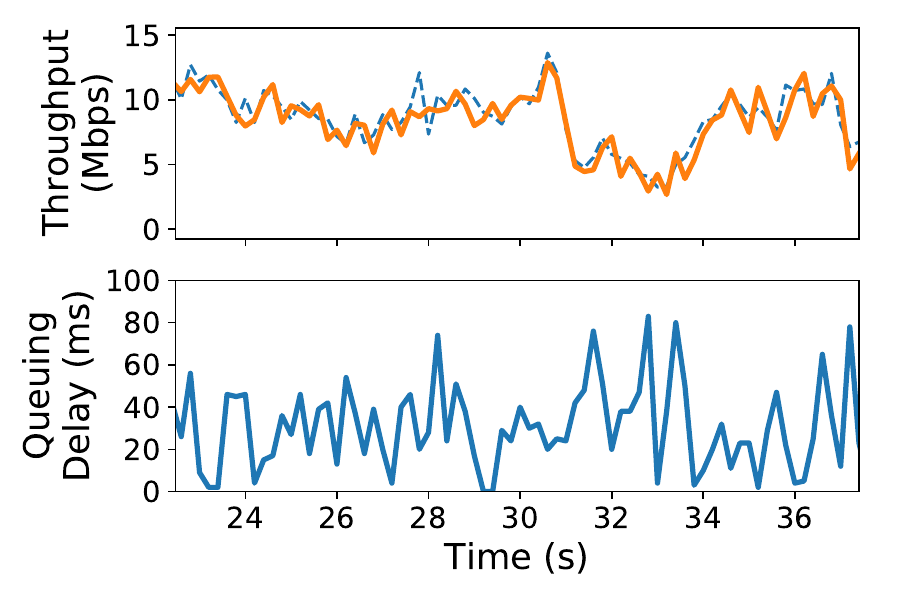}
        \vspace{-5mm}
    \caption{\small{\bf \name's robustness to flow size ---} With a single backlogged \name flow and multiple concurrent application-limited \name flows, all flows achieve high utilization and low delays.}
    \label{fig:traffic-char}
    \vspace{-7mm}
\end{figure}
We created a single long-lived \name flow that shared a cellular link with 200 application-limited \name flows that send traffic at an aggregate of 1 Mbit/s. \Fig{traffic-char} shows that, despite the fact that the application-limited flows do not have traffic to properly respond to \name's feedback, the \name flows (in aggregate) are still able to achieve low queuing delays and high link utilization.

\noindent\textbf{ABC's sensitivity to network latency:}

\begin{figure}[t]
     \centering
     \begin{subfigure}[t]{\columnwidth}
    \includegraphics[width=1.0\columnwidth]{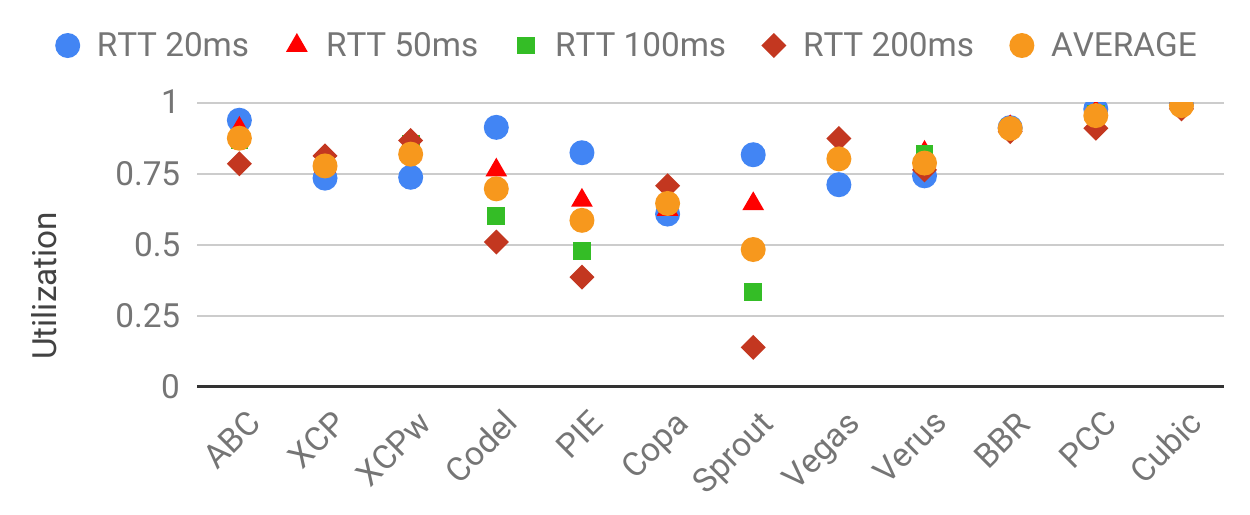}
    \vspace{-6mm}
    \subcaption{Utilization}
    \end{subfigure}
    \begin{subfigure}[t]{\columnwidth}
    \includegraphics[width=1.0\columnwidth]{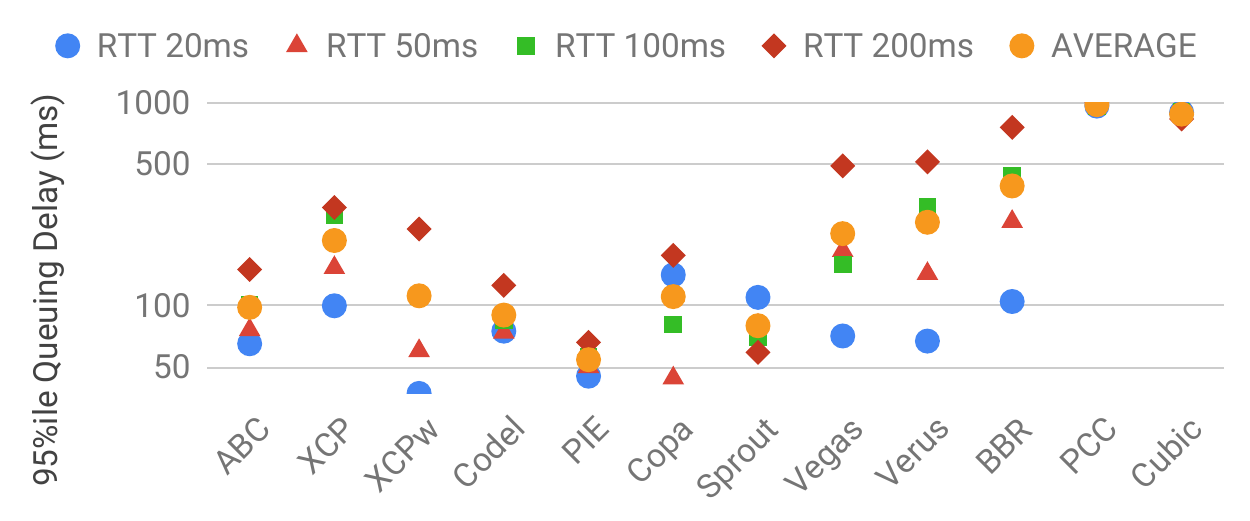} \vspace{-6mm}
    \subcaption{ per-packet queuing delay}
    \end{subfigure}
    \vspace{-3mm}
    \caption{\small {\bf Impact of propagation delay on performance ---} On a Verizon cellular network trace with different propagation delays, \name achieves a better throughput/delay tradeoff than all other schemes.}
    \label{fig:rtt}
    \vspace{-4mm}
 \end{figure}

Thus far, our emulation experiments have considered fixed minimum RTT values of 100 ms. To evaluate the impact that propagation delay has on \name's performance, we used a modified version of the experimental setup from \Fig{aggregate-statistics}. In particular, we consider RTT values of 20 ms, 50 ms, 100 ms, and 200 ms. \Fig{rtt} shows that, across all propagation delays, \name is still able to outperform all prior schemes, again achieving a more desirable throughput/latency trade off. \name's benefits persist even though schemes like Cubic+Codel and Cubic+PIE actually improve with decreasing propagation delays. Performance with these schemes improves because bandwidth delay products decrease, making Cubic's additive increase more aggressive (improving link utilization).

\end{document}